\documentclass[english]{article}
\usepackage[T1]{fontenc}
\usepackage[latin9]{inputenc}
\usepackage{rotating}
\usepackage{subfigure}
\usepackage{booktabs}
\usepackage{nicefrac}
\usepackage{graphicx}
\usepackage{amssymb}

\makeatletter

\providecommand{\tabularnewline}{\\}

\newcommand{\lyxaddress}[1]{
\par {\raggedright #1
\vspace{1.4em}
\noindent\par}
}

\makeatother

\usepackage{babel}

\begin{document}

\title{A Study of Elliptical Last Stable Orbits About a Massive Kerr Black
Hole}

\author{P. G. Komorowski{*}, S. R. Valluri{*}\#, M. Houde{*}}

\maketitle

\lyxaddress{{*}Department of Physics and Astronomy, \#Department of Applied Mathematics,
University of Western Ontario, London, Ontario}

pkomorow@uwo.ca, valluri@uwo.ca, houde@astro.uwo.ca 

\begin{abstract}
The last stable orbit (LSO) of a compact object (CO) is an important
boundary condition when performing numerical analysis of orbit evolution.
Although the LSO is already well understood for the case where a test-particle
is in an elliptical orbit around a Schwarzschild black hole (SBH)
and for the case of a circular orbit about a Kerr black hole (KBH)
of normalised spin, ${\tilde{S}}$ (${\left|{\mathbf{J}}\right|}/M^{2}$,
where ${\mathbf{J}}$ is the spin angular momentum of the KBH); it
is worthwhile to extend our knowledge to include elliptical orbits
about a KBH. This extension helps to lay the foundation for a better
understanding of gravitational wave (GW) emission. 

The mathematical developments described in this work sprang from the
use of an effective potential (${\tilde{V}}$) derived from the Kerr
metric, which encapsulates the Lense-Thirring precession. That allowed
us to develop a new form of analytical expression to calculate the
LSO Radius for circular orbits (${R_{LSO}}$) of arbitrary KBH spin.
We were then able to construct a numerical method to calculate the
latus rectum (${\tilde{l}}_{LSO}$) for an elliptical LSO. 

Formulae for ${\tilde{E}}^{2}$ (square of normalised orbital energy)
and ${\tilde{L}}^{2}$ (square of normalised orbital angular momentum)
in terms of eccentricity, ${\varepsilon}$, and latus rectum, ${\tilde{l}}$,
were previously developed by others for elliptical orbits around an
SBH and then extended to the KBH case; we used these results to generalise
our analytical ${{\tilde{l}}_{LSO}}$ equations to elliptical orbits.
LSO data calculated from our analytical equations and numerical procedures,
and those previously published, are then compared and found to be
in excellent agreement. 
\end{abstract}
04.20.-q, 95.30.Sf, 95.75.Pq

\section{Introduction}

One of the most important goals in experimental gravitation today
is the detection of gravitational wave (GW) radiation \cite{Thorne:1987fk,2002PhRvD..66d4002G,2005math.ph...3037C}.
To achieve this goal, considerable effort has been made to improve
the theoretical understanding of the evolution of compact object (CO)
orbits in extreme black hole systems \cite{PhysRev.131.435,PhysRev.136.B1224,1992PhRvD..46.5236F,1995PhRvD..52.3159R,Barack:2004uq,Flanigan:2007kx}.
In this paper, we assume a point-like test-particle since the ratio
of CO mass to the mass of the massive black hole (MBH) will be small
(i.e. $\leq{10}^{-5}$ \cite{2000PhRvD..62l4022O}); and the effect
of the CO mass upon the post Newtonian (PN) equations that we will
use in our subsequent modelling of the CO orbits is negligible \cite{1992MNRAS.254..146J,2002PhRvD..66d4002G,PhysRev.131.435}.
In this paper, we shall then assume the behaviour of the CO to be
closely approximated by that of a test-particle. If reference is made
to the orbital evolution of a true CO, as described by the PN evolution
equations, then we will use the term, CO. The treatment of CO orbital
evolution we will present in a forthcoming paper will be based on
the work of \cite{PhysRev.131.435,PhysRev.136.B1224,1991STIA...9240399B,1992MNRAS.254..146J,1992PhRvD..46.5236F,1995PhRvD..52.3159R,Barack:2004uq,2005PhRvL..94v1101H}
in which PN equations for a rotating MBH, also called a Kerr black
hole (KBH), are considered.

The objective of this study is to lay the foundation for our subsequent
work that will include the numerical calculation of the GW energy
emission by extreme KBH systems where the CO is in an elliptical orbit
in the equatorial plane of the KBH. The most basic quadrupole model
\cite{PhysRev.131.435,PhysRev.136.B1224} admitted solutions in closed
form \cite{PhysRev.136.B1224,1996NCimB.111..631P}; but because the
more comprehensive evolution equations now used are too complicated
to admit an analytical solution, numerical integration of orbital
parameters is required \cite{Barack:2004uq}. Therefore the last stable
orbit (LSO) becomes an important boundary condition. Such an analysis
requires an understanding of how ${\tilde{V}}$ depends on ${\tilde{L}}$
and on the inclination of the CO orbit. To undertake future work for
inclined orbits it is important to know the minimum physically meaningful
value of ${\tilde{L}}$. 

Previous research has demonstrated how the effective potential (${\tilde{V}}$)
of a test-particle in an equatorial orbit around a Schwarzschild black
hole (SBH) \cite{1983mtbh.book.....C,PhysRevD.50.3816,Schutz:2005fk}
can be calculated from the Schwarzschild metric and used to determine
the latus rectum of the LSO (${{\tilde{l}}_{LSO}}$). A treatment
of ${\tilde{V}}$ for a KBH system, where the test-particle follows
a circular LSO (section 12.7 of \cite{1983bhwd.book.....S}) yields
an analytical expression for the value ${R_{LSO}}$ in terms of normalised
spin, ${\tilde{S}}$, (equation 12.7.24 in \cite{1983bhwd.book.....S},
\cite{Bardeen:1972fi}) (${\tilde{S}}={\left|{\mathbf{s}}\right|}/M$
where ${\mathbf{s}}={\mathbf{J}}/M$ and ${\mathbf{J}}$ represents
the spin angular momentum of the KBH). Such treatment of ${\tilde{V}}$
also gives rise to expressions (equations 12.7.17 and 12.7.18 in \cite{1983bhwd.book.....S})
for the orbital energy, ${\tilde{E}}$, and orbital angular momentum,
${\tilde{L}}$. In \cite{PhysRevD.50.3816} the energy and orbital
angular momentum equations were also derived for an SBH system with
the test-particle in an elliptical orbit. In the significant work
by Glampedakis and Kennefick \cite{2002PhRvD..66d4002G}, their treatment
of ${\tilde{E}}$ and the quantity, $\left({\tilde{L}}-{\tilde{S}}{\tilde{E}}\right)$,
enabled us to derive generalised ${R_{LSO}}$ formulae for elliptical
orbits. Analytical expressions have a clear usefulness in the development
of new theoretical concepts and numerical methods \cite{2000PhRvD..62l4022O,2008PhRvD..77l4050S}.

The Lense-Thirring effect, an apodeictic \cite{2004Natur.431..958C}
prediction of general relativity, is the means by which the rotation
of the KBH imparts important changes on the test-particle orbit \cite{1984GReGr..16..711M,2000CQGra..17.2369C,2004Natur.431..958C,2007Natur.449...41C}
that are distinct from those associated with the SBH. The swirling
of spacetime in the vicinity of the KBH applies a torsion to the orbiting
test-particle; therefore, the orbit evolution will be altered, thus
causing changes in the point at which the test-particle reaches its
LSO. We shall develop an analytical and numerical methodology to calculate
the LSO of a test-particle in elliptical orbit about a KBH. Numerical
estimates of the latus rectum of the elliptical LSO orbits with respect
to KBH spin are available in the literature (Table I in \cite{2002PhRvD..66d4002G},
based upon the work of Schmidt \cite{2002CQGra..19.2743S}, and Table
I in \cite{2000PhRvD..62l4022O}); and they will provide a means to
validate our results. 

In section \ref{sub:Kerr-Metric}, the Kerr metric is introduced and
used in section \ref{sub:Effective-Potential} as the basis of developing
some essential analytical formulae to calculate the orbital angular
momentum of test-particles in circular paths around a KBH (section
\ref{sub:LastStableOrbit}). A formula for ${R}_{LSO}$ (prograde
and retrograde) is then developed analytically; and the general formula
for the ${{\tilde{l}}_{LSO}}$ of elliptical orbits is also presented.
In section \ref{sub:Calculating-the-LSO} the development and demonstration
of a numerical algorithm to determine the latus rectum and eccentricity
of test-particles of higher orbital angular momentum then follows.
The results of this analysis (section \ref{sub:Calculations}), as
well as results obtained from the general analytical formulae for
LSO latus rectum, are compared with results obtained from the literature.
In section \ref{sec:Conclusions} we shall draw conclusions.

\section{\label{sec:Understanding-the-Last}Understanding the Last Stable
Orbit About a Rotating Massive Black Hole}

\subsection{\label{sub:Kerr-Metric}Kerr Metric}

The Kerr metric (See equation 13.12 in \cite{2007PhT....60c..62H:Eq-13.12})
represents the solution to the Einstein Field Equations in the case
where the MBH possesses spin angular momentum,

\begin{equation}
g_{\alpha\beta}\Biggr|_{\mathit{Kerr}}=\left[\begin{array}{cccc}
-{\frac{\Delta-{M^{2}}{\tilde{S}}^{2}\sin^{2}\left(\theta\right)}{{\rho}^{2}}} & 0 & 0 & -2{M}{\frac{{M^{2}}\tilde{S}R\sin^{2}\left(\theta\right)}{{\rho}^{2}}}\\
\noalign{\medskip}0 & {\frac{{\rho}^{2}}{\Delta}} & 0 & 0\\
\noalign{\medskip}0 & 0 & {\rho}^{2} & 0\\
\noalign{\medskip}-2{M}{\frac{{M^{2}}\tilde{S}R\sin^{2}\left(\theta\right)}{{\rho}^{2}}} & 0 & 0 & {\frac{{M^{4}}\left({R}^{2}+{\tilde{S}}^{2}\right)^{2}-{M^{2}}{\tilde{S}}^{2}\Delta\,\sin^{2}\left(\theta\right)}{{\rho}^{2}}\sin^{2}\left(\theta\right)}\end{array}\right],\end{equation}
where $\rho={M}\sqrt{{R}^{2}+\cos^{2}\left(\theta\right){\tilde{S}}^{2}}$
and $\Delta={M^{2}}\left({R}^{2}-2\, R+{\tilde{S}}^{2}\right)$; in
which the factors, $R=r/M$ and $\tilde{S}={\left|\mathbf{J}\right|}/M^{2}$,
are used to express the metric in dimensionless terms. The symmetric
off-diagonal elements, $-2{M^{3}}\tilde{S}R\sin^{2}\left(\theta\right)/{\rho}^{2}$,
correspond to the Lense-Thirring precession that arises from the spin
of a central KBH of mass, ${M}$. Observe that when, $\tilde{S}=0$,
the Kerr metric equals the Schwarzschild Metric. 

Although the Schwarzschild Metric is expressed in spherical coordinates,
when the central black hole rotates it is appropriate to use the Kerr
metric expressed in Boyer-Lindquist (BL) coordinates. The conversion
of the BL coordinate system variables to Cartesian coordinate variables
is represented by these equations (see equation 11.4.7 in \cite{1990rcm..book.....D:Eq-11.4.7}
and see also \cite{1999CQGra..16.2929D}) :

\begin{eqnarray}
x & = & \sqrt{{R}_{BL}^{2}+{\tilde{S}}^{2}}\sin\left(\theta\right)\cos\left(\phi-f\right),\nonumber \\
y & = & \sqrt{{R}_{BL}^{2}+{\tilde{S}}^{2}}\sin\left(\theta\right)\sin\left(\phi-f\right),\nonumber \\
z & = & {R}_{BL}\cos\left(\theta\right),\label{eq:BL_coordinates}\end{eqnarray}
where \begin{equation}
{f}=\pm{\arctan\left({\nicefrac{\tilde{S}}{R_{BL}}}\right)}.\label{eq:arctanSR}\end{equation}
Because $0\leq{\tilde{S}}<1.0$, a prograde or retrograde orbit is
represented by the respective use of a plus or minus sign in equation
(\ref{eq:arctanSR}). The BL coordinate system will be used throughout
this treatment. The conversion of LSO radius from BL to spherical
coordinates is required whenever one performs a simulation of the
evolution equations reported in \cite{1993PhRvD..47.4183K,1995PhRvD..52.3159R,Ryan:1996ly,Hughes:2000dq,Barack:2004uq}.
This conversion is uncomplicated in the current application (in which
the angle, ${\iota}$, between the orbital angular momentum vector
and the spin axis of the KBH is zero), and proceeds by adding the
squares of $x,y,$ and $z$ as shown in equation (\ref{eq:BL_coordinates})
to obtain,\begin{equation}
{R}_{Spherical}^{2}={x}^{2}+{y}^{2}+{z}^{2}.\label{eq:BLxyz}\end{equation}
By substituting the relationships in equation (\ref{eq:BL_coordinates})
into equation (\ref{eq:BLxyz}) , one obtains the mathematical relationship,

\begin{equation}
R_{Spherical}^{2}=R_{BL}^{2}+\tilde{S}^{2}{\sin^{2}\left(\theta\right)}.\end{equation}
Recall that ${\tilde{S}}$ is the normalised spin of the KBH and ${\theta}$
is the polar angle of the test-particle in its orbit. In this study,
we work with orbits that are exclusively in the equatorial plane of
the KBH. Therefore one sets ${\theta}={\frac{\pi}{2}}$ to obtain

\begin{equation}
R_{Spherical}^{2}=R_{BL}^{2}+\tilde{S}^{2}.\label{eq:RSpherical-Equatorial}\end{equation}
Such a relationship is required for transforming LSO radii (BL coordinates)
into the spherical coordinate system.

\subsection{\label{sub:Effective-Potential}Effective Potential}

We shall develop a formulation of the effective potential of a test-particle
in orbit about a KBH. By so doing, the location of the LSO can be
estimated. In the following equations and calculations the radius,
${R}$, is represented in BL coordinates. For simplicity of notation,
the BL subscript will be suppressed (except in Section \ref{sub:Conversion-BL-Spherical}).

The four-momentum can be expressed as:\begin{eqnarray}
P_{\gamma} & = & \biggl[-E,\: m\frac{{\rho}^{2}}{\Delta}\left({\it \frac{dR}{d\tau}}\right),\:0,\:{m}{M}\tilde{L}\biggr]\label{eq:FourMomentum}\end{eqnarray}
for a particle of mass ${m}$ and,

\begin{eqnarray}
P_{\gamma} & = & \biggl[-E,\:\frac{{\rho}^{2}}{\Delta}\left({\it \frac{dR}{d\lambda}}\right),\:0,\: L\biggr],\end{eqnarray}
for zero mass, where $E$ is the energy of the orbital element and
$\tilde{L}$ is the orbital angular momentum of the particle in orbit
normalised with respect to its mass, $m$, and the KBH mass, ${M}$.
The $\left(dR/d\tau\right)$ is the derivative of the radial component
of the compact object with respect to the proper time, $\tau$. For
the zero-mass particle (which has no rest mass), ${L}$ is its total
linear momentum (\textit{vis.} ${L}={E}_{photon}/{c}$). The factor,
$dR/d\lambda$, is the derivative of the radial component of the zero-mass
particle with respect to an affine parameter, $\lambda$, which is
used in place of proper time, $\tau$, since a zero-mass particle
always follows a null path.

The invariant quantity of mass-energy can be calculated for each case
of a test-particle of infinitesimal mass \begin{eqnarray}
\vec{P}\cdot\vec{P} & = & \left.P_{\gamma}P_{\delta}g^{\delta\gamma}\right|_{Kerr}=-{m}^{2},\label{eq:PdotP_orbital_element}\end{eqnarray}
and a zero-mass particle

\begin{eqnarray}
\vec{P}\cdot\vec{P} & = & \left.P_{\gamma}P_{\delta}g^{\delta\gamma}\right|_{Kerr}=0.\end{eqnarray}
In that respect, the expected behaviour of a test mass will differ
from that of a zero-mass orbital element. From these equations, the
effective potential can be calculated by making a few assumptions
about the path taken by the orbiting zero-mass particle. The inverse
Kerr metric ($g^{\delta\gamma}$) is shown in Appendix A (equations
(\ref{eq:InverseKerrMetric}) and (\ref{eq:InverseKerrMetricThetaPiby2})).

\subsubsection{Test Particle.}

We restrict our work to the case of a test particle of mass, $m$,
in orbit about a KBH with ${\theta}={\frac{\pi}{2}}$. By evaluating
$\vec{P}\cdot\vec{P}$ (see equation (\ref{eq:PdotP_orbital_element}))
using the test mass four-momentum (see equation (\ref{eq:FourMomentum}))
one obtains, 

\begin{eqnarray}
\vec{P}\cdot\vec{P} & =- & \Biggl({R}^{4}{E}^{2}-{R}^{4}{m}^{2}\left(\frac{dR}{d\tau}\right)^{2}-{R}^{2}{m}^{2}{\tilde{L}}^{2}+{R}^{2}{E}^{2}{\tilde{S}}^{2}\nonumber \\
 &  & +2\, R{E}^{2}{\tilde{S}}^{2}+2\, R{m}^{2}{\tilde{L}}^{2}+4\, RE\tilde{S}m\tilde{L}\Biggr)\nonumber \\
 & \times & \left({R}^{4}-2\,{R}^{3}+{R}^{2}{\tilde{S}}^{2}\right)^{-1}\nonumber \\
 & = & -m^{2}.\label{eq:PdotPm2}\end{eqnarray}

To develop a relationship between the effective potential and the
orbital parameters several sequential steps must be followed. First,
all terms in equation (\ref{eq:PdotPm2}) are collected and equated
to zero, and then divided by $m^{2}$ and the $\left({dR}/{d\tau}\right)^{2}$
terms are collected on the right hand side of the equation. Noting
that ${E/m}={\tilde{E}}$ represents the specific energy content of
the orbiting test-particle, one then obtains,

\begin{eqnarray}
 &  & \left({R^{2}{\tilde{S}}^{2}+2R{\tilde{S}}^{2}+R^{4}}\right)\tilde{E}^{2}-\left(4R{\tilde{S}}\tilde{L}\right)\tilde{E}\nonumber \\
 & - & \left(\tilde{L}^{2}R^{2}-2\tilde{L}^{2}R+R^{2}\left({R^{2}-2R+{\tilde{S}}^{2}}\right)\right)\nonumber \\
 & = & R^{4}\left({\frac{{dR}}{{d\tau}}}\right)^{2}\end{eqnarray}

\noindent \begin{flushleft}
At the points of closest (pericentre) and farthest (apocentre) approach
the derivative of ${R}$ with respect to $\tau$ is zero. By performing
that simplification, one obtains a quadratic equation in $\tilde{E}$,
i.e.\begin{eqnarray}
 & - & \left({R^{2}{\tilde{S}}^{2}+2R{\tilde{S}}^{2}+R^{4}}\right)\tilde{E}^{2}+\left(4R{\tilde{S}}\tilde{L}\right)\tilde{E}\nonumber \\
 & + & \left(\tilde{L}^{2}R^{2}-2\tilde{L}^{2}R+R^{2}\left({R^{2}-2R+{\tilde{S}}^{2}}\right)\right)\nonumber \\
 & = & 0.\label{eq:Quadratic_der}\end{eqnarray}

\par\end{flushleft}

\noindent \begin{flushleft}
The factored form of equation (\ref{eq:Quadratic_der}) corresponds
to the following equation \cite{Schutz:2005fk}: \begin{eqnarray}
\left(\tilde{E}-{\tilde{V}}_{+}\right)\left(\tilde{E}-{\tilde{V}}_{-}\right) & = & 0.\end{eqnarray}
Therefore two solutions for the effective potential can be calculated:
\par\end{flushleft}

\begin{sidewaysfigure}
\includegraphics[scale=0.45]{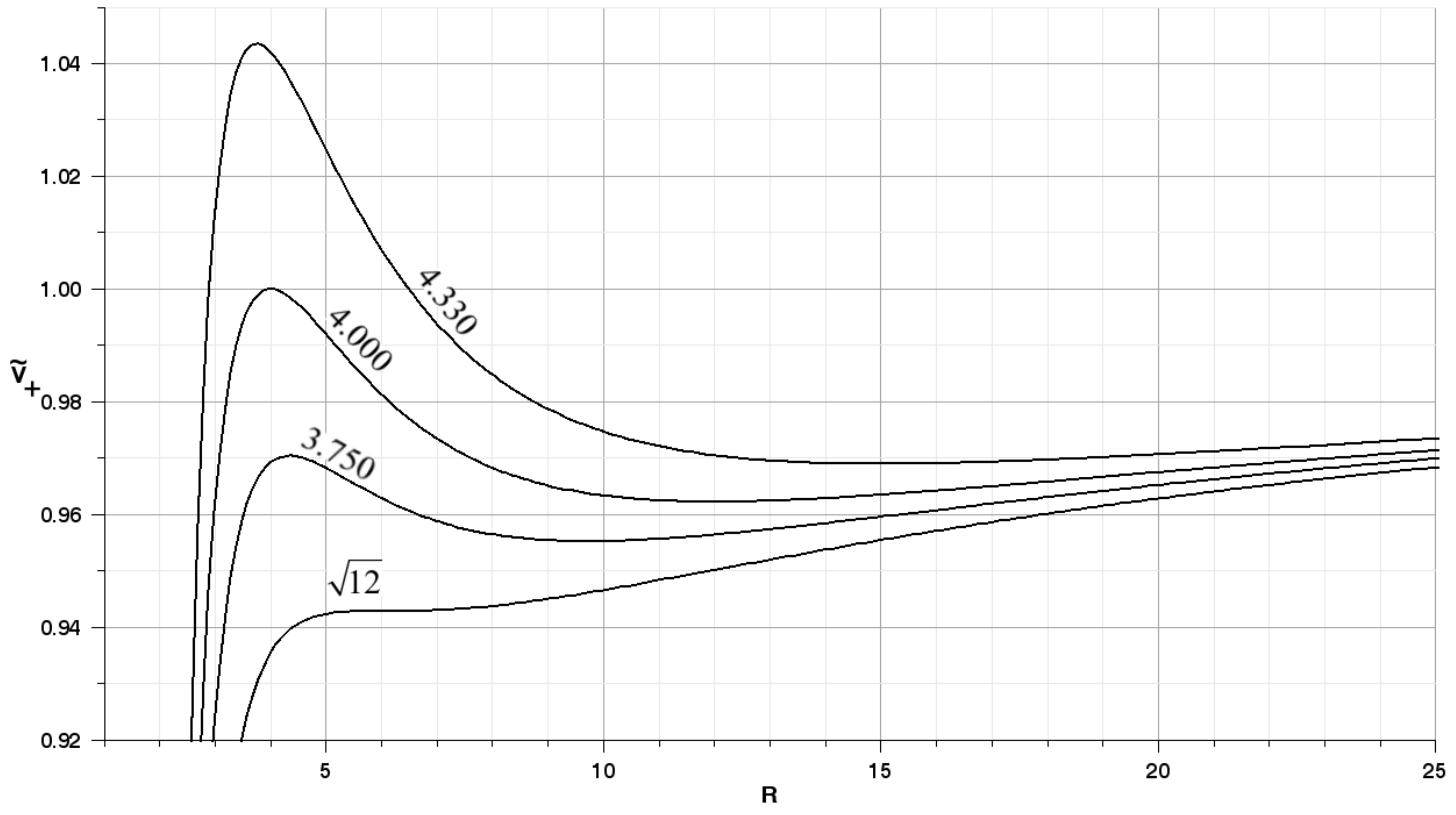}\caption{\label{fig:Potentials_For_Different_L}Effective Potentials for various
values of ${\tilde{L}}$ where ${\tilde{S}}=0$.}

\end{sidewaysfigure}

\noindent \begin{flushleft}
\begin{eqnarray}
\tilde{V}_{\pm} & = & \frac{-b\mp\sqrt{b^{2}-4ac}}{2a}\label{eq:PotentialPlusMinus}\\
a & = & -\left({R}^{4}+{R}^{2}{\tilde{S}}^{2}+2\,{\tilde{S}}^{2}R\right)\nonumber \\
b & = & 4\, R\tilde{S}\tilde{L}\nonumber \\
c & = & \left(L^{2}R^{2}-2L^{2}R+R^{2}\left({R^{2}-2R+{\tilde{S}}^{2}}\right)\right).\nonumber \end{eqnarray}
For the SBH (i.e. ${\tilde{S}}=0$), the value of ${\tilde{V}_{\pm}}^{2}$
(from equation (\ref{eq:PotentialPlusMinus})) becomes:
\par\end{flushleft}

\begin{eqnarray}
{\tilde{V}_{\pm}}^{2} & = & \frac{\left(R-2\right)\left({R}^{2}+{\tilde{L}}^{2}\right)}{{R}^{3}},\label{eq:Schwartzschild_Limit}\end{eqnarray}
which depends only on the values of $R$ and $\tilde{L}$, as expected
(as shown in Figure \ref{fig:Potentials_For_Different_L}).

The effective potential contains important information. In the case
of the SBH, the relationship between ${V}_{\pm}$ and ${R}$ describes
the test-particle orbit and leads us to a calculation of the values
of $\tilde{L}$ and ${R}$ at which the test-particle can no longer
sustain a stable orbit. The LSO is an important characteristic of
the binary system that is identified as the point at which the ${\tilde{V}}_{+}$
curve (Figure \ref{fig:Potentials_For_Different_L}) has a slope of
zero and the second derivative with respect to ${R}$ is not positive.
The effective potential, ${\tilde{V}}_{-}$, corresponds to particles
and photons for which their orbital angular momentum has an opposite
sense to the KBH spin{\large{} (section 11.3 in \cite{Schutz:2005fk}}).
The mathematical treatment of ${\tilde{V}}_{+}$ presented in the
sections that follow preserves its prograde and retrograde properties;
indeed, we have found that the use of ${\tilde{V}}_{-}$ in the calculations
that follow yield the same results.

\subsection{\label{sub:LastStableOrbit}Last Stable Orbit (LSO) for a CO in the
Equatorial Plane of the Kerr Black Hole}

\noindent The equations for the radius of a circular or elliptical
LSO can be calculated through a mathematical treatment of the following
two equations:\begin{eqnarray}
\frac{d{\tilde{V}}_{+}}{{dR}} & = & 0\label{eq:dVdR-locus}\end{eqnarray}
and

\begin{eqnarray}
\frac{d^{2}{\tilde{V}}_{+}}{{dR}^{2}} & \leq & 0,\label{eq:d2VdR2-locus}\end{eqnarray}
where the point of inflection (which corresponds to a circular LSO)
can be found by evaluating the intersection points of the equations
(\ref{eq:dVdR-locus}) and (\ref{eq:d2VdR2-locus}). 

\begin{figure}
\includegraphics[angle=90,scale=0.5]{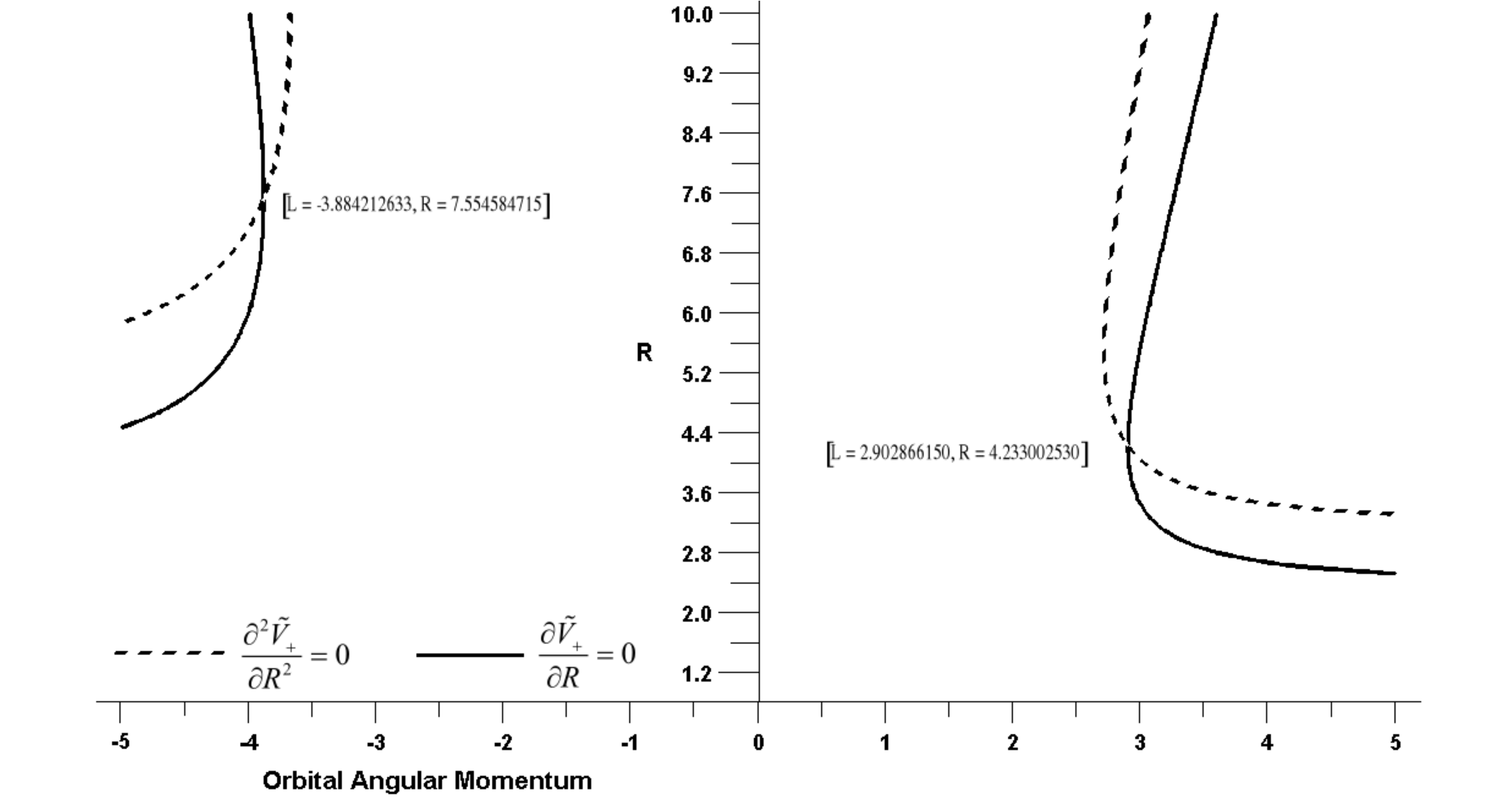}\caption{A plot of ${R}_{LSO}$ vs. ${\tilde{L}}$ for the first and second
derivatives of ${\tilde{V}}$ with respect to ${R}$.}
\label{fig:RvsL}
\end{figure}

The loci of these two equations is depicted in the $\left({R,\tilde{L}}\right)$
plane for a KBH with a spin value of ${\tilde{S}}=0.5$ (see Figure
\ref{fig:RvsL}). Their intersection points (derived numerically with
Maple 11), $\left[\begin{array}{cc}
{R}=7.554584715, & {\tilde{L}}=-3.884212633\end{array}\right]$ and $\left[\begin{array}{cc}
{R}=4.233002530, & {\tilde{L}}=2.902866150\end{array}\right]$, correspond to the radial position of the LSO, $R$, of a test-particle
with an orbital angular momentum of ${\tilde{L}}$. These points differ
from $\left[\begin{array}{cc}
{R}=6.0, & {\tilde{L}}=\pm\sqrt{12}\end{array}\right]$, which is the solution for an SBH. The existence of an intersection
point on the graphical plot notwithstanding (see Figure \ref{fig:RvsL}),
on frequent occasions, no result was returned by Maple. On other occasions
a correct value of ${R}$ was returned, while the value calculated
for ${\tilde{L}}$ deviated by at least a factor of two from the graphical
result. Such inconsistent behaviour was attributed to the great complexity
of the expressions being treated and the associated floating point
round off error; therefore, an analytic method was sought.

\subsubsection{Orbital Angular Momentum.}

The derivative of ${\tilde{V}_{+}}$, equated to zero, can be used
to determine an analytical expression for $\tilde{L}^{2}$ in terms
of $R$ and ${\tilde{S}}$ for circular or elliptical orbits. From
equations (\ref{eq:PotentialPlusMinus}) and (\ref{eq:dVdR-locus})
one obtains,

\begin{eqnarray}
\frac{d\tilde{V}_{+}}{dR} & = & \Biggl(-3\,{S}^{2}{R}^{4}{L}^{2}+6\,{S}^{2}{R}^{3}{L}^{2}-2\,{R}^{2}{S}^{4}{L}^{2}\nonumber \\
 &  & +{R}^{5}{L}^{2}{S}^{2}+2\,{S}^{2}{R}^{5}-3\,{R}^{6}{S}^{2}-3\,{R}^{6}{L}^{2}\nonumber \\
 &  & -3\,{S}^{4}{R}^{4}-{S}^{6}{R}^{2}-2\, R{S}^{6}+8\,{S}^{4}{R}^{2}-{R}^{8}+{R}^{7}{L}^{2}\nonumber \\
 &  & +\left(6\,{R}^{2}SL+2\,{S}^{3}L\right)\sqrt{{R}^{3}\left({R}^{2}-2\, R+{S}^{2}\right)\left({R}^{3}+{L}^{2}R+{S}^{2}R+2\,{S}^{2}\right)}\Biggr)\nonumber \\
 & \div & \Biggl(\sqrt{{R}^{3}\left({R}^{2}-2\, R+{S}^{2}\right)\left({R}^{3}+{L}^{2}R+{S}^{2}R+2\,{S}^{2}\right)}\nonumber \\
 &  & \left({R}^{3}+{S}^{2}R+2\,{S}^{2}\right)^{2}\Biggr)\nonumber \\
 & = & 0.\label{eq:dVdR-0}\end{eqnarray}
The denominator of equation (\ref{eq:dVdR-0}) can be disregarded
because the quotient is equated to zero; it is also required that
the roots of the factors present in the denominator lie outside the
range of physically attainable ${R}$ values. To be specific, the
roots of $\left({R}^{2}-2\, R+{\tilde{S}}^{2}\right)$ correspond
to the event horizon for massless particles, those of ${R}^{3}$ are
zero and beyond the LSO, and the roots of $\left({R}^{3}+{\tilde{L}}^{2}R+{\tilde{S}}^{2}R+2\,{\tilde{S}}^{2}\right)$
and $\left({R}^{3}+{\tilde{S}}^{2}R+2\,{\tilde{S}}^{2}\right)$ are
complex and thus also physically unattainable for real values of ${R}$. 

The simplified power series is thus derived from the numerator of
equation (\ref{eq:dVdR-0}) after eliminating the square root,

\begin{eqnarray}
 &  & {R}^{3}\left(9\, R-6\,{R}^{2}+{R}^{3}-4\,{\tilde{S}}^{2}\right){\tilde{L}}^{4}\nonumber \\
 &  & -2\,{R}^{2}\left(-3\,{R}^{4}+{R}^{5}-12\,{\tilde{S}}^{2}R+6\,{R}^{2}{\tilde{S}}^{2}+2\,{R}^{3}{\tilde{S}}^{2}+5\,{\tilde{S}}^{4}+{\tilde{S}}^{4}R\right){\tilde{L}}^{2}\nonumber \\
 &  & +\left({R}^{4}+2\,{R}^{2}{\tilde{S}}^{2}-4\,{\tilde{S}}^{2}R+{\tilde{S}}^{4}\right)^{2}\nonumber \\
 &  & =0.\end{eqnarray}
Therefore $\tilde{L}^{2}$ can be obtained directly by using the quadratic
formula, 

\begin{equation}
\tilde{L}^{2}=\frac{-b\pm\sqrt{b^{2}-4ac}}{2a},\label{eq:L2BAC}\end{equation}
where we have redefined:\begin{eqnarray*}
{a} & = & {R}^{3}\left(9\, R-6\,{R}^{2}+{R}^{3}-4\,{\tilde{S}}^{2}\right)\\
{b} & = & -2\,{R}^{2}\left(-3\,{R}^{4}+{R}^{5}-12\,{\tilde{S}}^{2}R+6\,{R}^{2}{\tilde{S}}^{2}+2\,{R}^{3}{\tilde{S}}^{2}+5\,{\tilde{S}}^{4}+{\tilde{S}}^{4}R\right)\\
{c} & = & \left({R}^{4}+2\,{R}^{2}{\tilde{S}}^{2}-4\,{\tilde{S}}^{2}R+{\tilde{S}}^{4}\right)^{2}.\end{eqnarray*}
Two solutions are found that correspond to the orbital angular momenta
of a test-particle in a prograde orbit,

\begin{eqnarray}
{\tilde{L}}_{Pro}^{2} & = & \left(-3\,{R}^{6}+{R}^{7}-12\,{R}^{3}{\tilde{S}}^{2}+6\,{R}^{4}{\tilde{S}}^{2}\right.\nonumber \\
 &  & +2\,{R}^{5}{\tilde{S}}^{2}+5\,{\tilde{S}}^{4}{R}^{2}+{\tilde{S}}^{4}{R}^{3}\nonumber \\
 &  & \left.-2\,{\tilde{S}}\left(3\,{R}^{2}+{\tilde{S}}^{2}\right)\left({R}^{2}-2\, R+{\tilde{S}}^{2}\right)\sqrt{{R}^{3}}\right)\nonumber \\
 &  & \left({R}^{3}\left(9\, R-6\,{R}^{2}+{R}^{3}-4\,{\tilde{S}}^{2}\right)\right)^{-1},\label{eq:L2-Quadratic-Solution-Pro}\end{eqnarray}
and in a retrograde orbit,

\begin{eqnarray}
{\tilde{L}}_{Ret}^{2} & = & \left(-3\,{R}^{6}+{R}^{7}-12\,{R}^{3}{\tilde{S}}^{2}+6\,{R}^{4}{\tilde{S}}^{2}\right.\nonumber \\
 &  & +2\,{R}^{5}{\tilde{S}}^{2}+5\,{\tilde{S}}^{4}{R}^{2}+{\tilde{S}}^{4}{R}^{3}\nonumber \\
 &  & \left.+2\,{\tilde{S}}\left(3\,{R}^{2}+{\tilde{S}}^{2}\right)\left({R}^{2}-2\, R+{\tilde{S}}^{2}\right)\sqrt{{R}^{3}}\right)\nonumber \\
 &  & \left({R}^{3}\left(9\, R-6\,{R}^{2}+{R}^{3}-4\,{\tilde{S}}^{2}\right)\right)^{-1}.\label{eq:L2-Quadratic-Solution-Retro}\end{eqnarray}

An analytical expression for ${\tilde{L}}^{2}$ with respect to ${R}$
and ${\tilde{S}}$ has been derived. But one must consider that the
formula is limited to providing a value of ${\tilde{L}}^{2}$ that
corresponds to a test-particle in its LSO (BL coordinates) about a
KBH of spin ${\tilde{S}}$. These formulae (equations (\ref{eq:L2-Quadratic-Solution-Pro})
and (\ref{eq:L2-Quadratic-Solution-Retro})) do not provide a relationship
between ${R}$ and ${\tilde{L}}^{2}$ for a general orbit.

Consider an example where ${\tilde{S}}=0.5$. The relationship between
the value of ${\tilde{L}}$ and the radius $R\in[1.0,\:6.0]$ is plotted
in Figure \ref{fig:Min-of-L-wrt-R}. One observes a power series for
which the values of ${R}$ and ${\tilde{L}}$ for a circular orbit
(at the point of inflection) occur at the local minimum. Therefore
it is possible to derive an expression for the radius of the LSO,
${R}_{LSO}$, at that point of inflection for an arbitrary spin, ${\tilde{S}}$,
where $0\leq{\tilde{S}}<1$.

\subsubsection{Circular LSO Radius.}

The calculation of such an analytical relationship proceeds as follows.
The derivative of ${\tilde{L}}^{2}$ with respect to ${R}$ is set
equal to zero. From equation (\ref{eq:L2BAC}) we obtain:

\begin{eqnarray}
\frac{d\left({\tilde{L}}^{2}\right)}{d{R}} & = & \Biggl[{\tilde{S}}\sqrt{{R}}/{R}\left(3R^{7}-45R^{5}+20{\tilde{S}}^{2}R^{5}+54R^{4}-26{\tilde{S}}^{2}R^{4}\right.\nonumber \\
 &  & +9{\tilde{S}}^{4}R^{3}+24{\tilde{S}}^{2}R^{3}-26{\tilde{S}}^{4}R^{2}-54{\tilde{S}}^{2}R^{2}+53{\tilde{S}}^{4}R-12{\tilde{S}}^{6}\Bigr)\nonumber \\
 & \pm & \Bigl(R^{8}-2R^{6}{\tilde{S}}^{2}-3R^{4}{\tilde{S}}^{4}-12R^{7}-28{\tilde{S}}^{2}R^{5}-24{\tilde{S}}^{4}R^{3}\nonumber \\
 &  & +45R^{6}+126{\tilde{S}}^{2}R^{4}+57{\tilde{S}}^{4}R^{2}+20{\tilde{S}}^{6}\nonumber \\
 &  & -54R^{5}-144{\tilde{S}}^{2}R^{3}-90{\tilde{S}}^{4}R+108{\tilde{S}}^{2}R^{2}\Bigr)\Biggr]\nonumber \\
 & \times & \left[{R^{2}\left({-R^{3}+4{\tilde{S}}^{2}+6R^{2}-9R}\right)^{2}}\right]^{-1}=0.\end{eqnarray}
Where the plus sign corresponds to a prograde orbit and the minus
sign corresponds to a retrograde orbit. The denominator contains a
factor (i.e. $\left({-R^{3}+4{\tilde{S}}^{2}+6R^{2}-9R}\right)$)
with roots that correspond to the event horizon and a factor $R^{2}$
with roots equal to zero, which lie beyond the event horizon and are
thus unattainable.

{\small }%
\begin{sidewaystable}
{\small \caption{\label{tab:Summary-of-Factors}A summary of the factors found in equation
(\ref{eq:Poly}) for ${\tilde{S}}=0.0,\:0.5,$ and $1.0$.}
}{\small \par}

{\small }\begin{tabular}{lcccc}
\hline 
{\small Factor} & {\small ${\tilde{S}}=0.0$} & {\small ${\tilde{S}}=0.5$} & {\small ${\tilde{S}}=1.0$} & {\small Comments}\tabularnewline
\hline
{\small $\left({\tilde{S}}^{2}-{R}^{3}\right)$} & {\small $\left\{ \begin{array}{c}
0\\
0\\
0\end{array}\right\} $} & {\small $\left\{ \begin{array}{c}
0.629961\\
-0.315\left(1\pm i\,\sqrt{3}\right)\end{array}\right\} $} & {\small $\left\{ \begin{array}{c}
1.0\\
-0.5\left(1\pm i\,\sqrt{3}\right)\end{array}\right\} $} & \tabularnewline
 &  &  &  & \tabularnewline
{\small $\begin{array}{c}
\left(9\,{\tilde{S}}^{4}-28\,{\tilde{S}}^{2}R-6\,{\tilde{S}}^{2}{R}^{2}\right.\\
\left.+36\,{R}^{2}-12\,{R}^{3}+{R}^{4}\right)\end{array}$} & {\small $\left\{ \begin{array}{c}
0\\
0\\
6\\
6\end{array}\right\} $} & {\small $\left\{ \begin{array}{c}
0.10620\ldots\pm i\,0.079436\ldots\\
4.233002530\\
7.554584715\end{array}\right\} $} & {\small $\left\{ \begin{array}{c}
1\\
1\\
1\\
9\end{array}\right\} $} & {\small }\begin{tabular}{l}
{\small Corresponds to radius of LSO}\tabularnewline
{\small for test mass particles.}\tabularnewline
\tabularnewline
\end{tabular}\tabularnewline
 &  &  &  & \tabularnewline
{\small $\left({\tilde{S}}^{4}+2\,{\tilde{S}}^{2}{R}^{2}-4\,{\tilde{S}}^{2}R+{R}^{4}\right)$} & {\small $\left\{ \begin{array}{c}
0\\
0\\
0\\
0\end{array}\right\} $} & {\small $\left\{ \begin{array}{c}
0.06460427\ldots\\
0.80423269\ldots\\
-0.4344\pm i\,1.00708\ldots\end{array}\right\} $} & {\small $\left\{ \begin{array}{c}
0.295598\ldots\\
1\\
-0.64780\pm i\,1.7214\ldots\end{array}\right\} $} & \tabularnewline
 &  &  &  & \tabularnewline
{\small $\left({R}^{3}-4\,{\tilde{S}}^{2}-6\,{R}^{2}+9\, R\right)$} & {\small $\left\{ \begin{array}{c}
0\\
3\\
3\end{array}\right\} $} & {\small $\left\{ \begin{array}{c}
0.120615\ldots\\
2.347300\ldots\\
3.532090\ldots\end{array}\right\} $} & {\small $\left\{ \begin{array}{c}
1\\
1\\
4\end{array}\right\} $} & {\small }\begin{tabular}{l}
{\small Corresponds to radius of LSO}\tabularnewline
{\small for massless particles}\tabularnewline
{\small (e.g. photons or gravitons).}\tabularnewline
\end{tabular}\tabularnewline
\end{tabular}
\end{sidewaystable}
{\small \par}

The simplification of the equation by taking only the numerator and
eliminating the square root, can proceed to yield the following result:

\begin{eqnarray}
 &  & {R}^{3}\left({\tilde{S}}^{2}-{R}^{3}\right)\left(9\,{\tilde{S}}^{4}-28\,{\tilde{S}}^{2}R-6\,{\tilde{S}}^{2}{R}^{2}+36\,{R}^{2}-12\,{R}^{3}+{R}^{4}\right)\nonumber \\
 & \times & \left({\tilde{S}}^{4}+2\,{\tilde{S}}^{2}{R}^{2}-4\,{\tilde{S}}^{2}R+{R}^{4}\right)\left({R}^{3}-4\,{\tilde{S}}^{2}-6\,{R}^{2}+9\, R\right)^{2}\nonumber \\
 & = & 0.\label{eq:Poly}\end{eqnarray}

Fortunately, the polynomial that expresses the relationship between
${\tilde{S}}$ and ${R}_{LSO}$, is already simplified into a product
of some binomials, trinomials, and quartics (see Table \ref{tab:Summary-of-Factors}).
Each one can be assessed by considering the examples of an SBH with
no spin (${\tilde{S}}=0.0$) and a KBH with ${\tilde{S}}=0.5$. For
the former case, the solution, $\left[\begin{array}{cc}
{R}=6.0, & {\tilde{L}}=\sqrt{12}\end{array}\right]$, is known; for the second case, it has been calculated numerically,
$\left[\begin{array}{cc}
{R}=4.233002530, & {\tilde{L}}=2.902866150\end{array}\right]$. These cases help one to identify the relevant factor. It is interesting
to observe that some of the radii in Table \ref{tab:Summary-of-Factors}
have complex values.

The factor that yields the values of the LSO radii (one for each of
the possible prograde and retrograde orbits of the CO) is:

\begin{equation}
\left(9\,{\tilde{S}}^{4}-28\,{\tilde{S}}^{2}R-6\,{\tilde{S}}^{2}{R}^{2}+36\,{R}^{2}-12\,{R}^{3}+{R}^{4}\right)=0.\label{eq:Quartic-important}\end{equation}
This quartic equation (\ref{eq:Quartic-important}) can be converted
to a companion matrix which is solved for its eigenvalues to yield
the analytical solutions for ${R}_{LSO}$ for the prograde and retrograde
orbits (see Appendix B). These solutions are:

\begin{eqnarray}
{R}_{pro} & = & 3+\sqrt{{Z}}-\sqrt{\frac{16{\tilde{S}}^{2}}{\sqrt{{Z}}}-{Z}+3\left(3+{\tilde{S}}^{2}\right)}\label{eq:RLSO_Prograde}\end{eqnarray}
and\begin{eqnarray}
{R}_{ret} & = & 3+\sqrt{{Z}}+\sqrt{\frac{16{\tilde{S}}^{2}}{\sqrt{{Z}}}-{Z}+3\left(3+{\tilde{S}}^{2}\right)}\label{eq:RLSO_Retrograde}\end{eqnarray}
where:\begin{eqnarray*}
Z & = & 3+{\tilde{S}}^{2}+(3+\tilde{S})\left((1+\tilde{S})(1-\tilde{S})^{2}\right)^{\frac{1}{3}}\\
 &  & +(3-\tilde{S})\left((1-\tilde{S})(1+\tilde{S})^{2}\right)^{\frac{1}{3}}.\end{eqnarray*}

Although formulae that are analytically the same as ours have already
been developed by Bardeen et al. \cite{Bardeen:1972fi}, our formulae
were derived by independent means and are simpler. The numerical results
of each equation differ insignificantly over the physically valid
range of $0\leq{\tilde{S}}<1.0$. And our formulae are more robust
with respect to round-off error when evaluated numerically, and they
are roborant of the pre-existing calculations.

\subsubsection{Orbital Energy and Angular Momentum at the LSO.}

One can derive new formulae for the test-particle orbital energy,
${\tilde{E}}$, and angular momentum, ${\tilde{L}}$, in terms of
parameters ${\tilde{S}}$, ${\varepsilon}$, and latus rectum, ${\tilde{l}}$,
by using equation (\ref{eq:PotentialPlusMinus}) as a starting point.
We know that, \begin{equation}
{\tilde{V}}_{+}={\tilde{E}},\end{equation}
$\Rightarrow$

\begin{eqnarray}
{\tilde{E}} & = & \Biggl[{2R{\tilde{S}}\tilde{L}+}\nonumber \\
 &  & \sqrt{R\left({R^{2}-2R+{\tilde{S}}^{2}}\right)\left({R^{5}+R^{3}\tilde{L}^{2}+R^{3}{\tilde{S}}^{2}+2R^{2}{\tilde{S}}^{2}}\right)}\Biggr]\nonumber \\
 &  & \Biggl[{R\left({R^{3}+{\tilde{S}}^{2}R+2{\tilde{S}}^{2}}\right)}\Biggr]^{-1}\label{eq:ERSL}\end{eqnarray}
when the test-particle is in its LSO. Although the roots in ${R}$
are readily found by Maple, they are inordinately long and not useful.
A more effective derivation method, similar to the one used in \cite{2002PhRvD..66d4002G},
shall be outlined.

By manipulating the formula in equation (\ref{eq:ERSL}) we obtain:\begin{eqnarray}
{R}^{3}-\left(\frac{2}{1-{\tilde{E}}^{2}}\right){R}^{2}+\left(\frac{{\tilde{L}}^{2}+{\tilde{S}}^{2}-{\tilde{E}}^{2}{\tilde{S}}^{2}}{1-{\tilde{E}}^{2}}\right){R}\nonumber \\
-\left(\frac{2\left({\tilde{L-}\tilde{E}\tilde{S}}\right)^{2}}{1-{\tilde{E}}^{2}}\right) & = & 0\end{eqnarray}
from which we can obtain the expressions for the sum and the product
of the roots in ${R}$ directly from the coefficients of the polynomial,
\textit{vis.} \begin{eqnarray}
\left(R-r_{1}\right)\left(R-r_{2}\right)\left(R-r_{3}\right) & = & 0\label{eq:Poly1}\end{eqnarray}
which implies,\begin{eqnarray}
R^{3}-\left({r_{1}+r_{2}+r_{3}}\right)R^{2}+\left({r_{1}r_{3}+r_{1}r_{2}+r_{2}r_{3}}\right)R-r_{1}r_{2}r_{3} & = & 0.\label{eq:Poly2}\end{eqnarray}
Where ${\left\{ r_{1},r_{2},r_{3}\right\} }$ are the roots in (\ref{eq:Poly1})
and (\ref{eq:Poly2}). We find the following equations for the sum
of the ${R}$ roots (i.e. ${R}_{sum}=r_{1}+r_{2}+r_{3}$):

\begin{equation}
R_{sum}=2\left({1-\tilde{E}^{2}}\right)^{-1},\end{equation}
and for their product (i.e. ${R}_{prod}=r_{1}r_{2}r_{3}$),

\begin{equation}
R_{prod}=R_{sum}\left({\tilde{L}-\tilde{E}\tilde{S}}\right)^{2}.\end{equation}
The corresponding formulae for ${\tilde{E}}$ and ${\tilde{L}}$ are
as follows:

\begin{equation}
\tilde{E}=\pm\frac{{\sqrt{R_{sum}\left({R_{sum}-2}\right)}}}{{R_{sum}}},\label{eq:E-Rsum}\end{equation}
and,

\begin{equation}
\tilde{L}=\frac{{\sqrt{R_{sum}\left({R_{sum}-2}\right)}{\tilde{S}}\pm\sqrt{R_{sum}R_{prod}}}}{{R_{sum}}}.\label{eq:L-Rsum}\end{equation}
For the LSO, the roots, ${\left\{ r_{1},r_{3}\right\} }$, correspond
to the LSO radius; therefore, we make the following substitutions:

\begin{eqnarray}
{r}_{1}={r}_{3}=R_{Min} & = & \frac{\tilde{l}}{{1+\varepsilon}},\end{eqnarray}
and\begin{eqnarray}
{r}_{2}=R_{Max} & = & \frac{\tilde{l}}{{1-\varepsilon}},\end{eqnarray}
where ${\tilde{l}}$ is the latus rectum of the elliptical LSO. We
can now set: 

\begin{eqnarray}
R_{sum} & = & 2R_{Min}+R_{Max}\nonumber \\
 & = & 2\frac{\tilde{l}}{{1+\varepsilon}}+\frac{\tilde{l}}{{1-\varepsilon}}\label{eq:Rsum}\end{eqnarray}
and

\begin{eqnarray}
R_{Prod} & = & {R_{Min}}^{2}R_{Max}\nonumber \\
 & = & \frac{{{\tilde{l}}^{3}}}{{\left({1+\varepsilon}\right)^{2}\left({1-\varepsilon}\right)}}.\label{eq:Rprod}\end{eqnarray}

By substituting equations (\ref{eq:Rsum}) and (\ref{eq:Rprod}) into
equations (\ref{eq:E-Rsum}) and (\ref{eq:L-Rsum}) the following
formulae are obtained:

\begin{eqnarray}
{\tilde{E}}^{2} & = & {1-2\left({1-\varepsilon^{2}}\right)\left({\tilde{l}}\left({3-\varepsilon}\right)\right)^{-1}}\label{eq:MyE2}\end{eqnarray}
and,\begin{eqnarray}
\tilde{L}^{2} & = & \left({{\tilde{S}}{\tilde{E}}\pm{\tilde{l}}\sqrt{\frac{1}{\left({1+\varepsilon}\right)\left({3-\varepsilon}\right)}}}\right)^{2}\label{eq:MyL2SE}\end{eqnarray}
$\Rightarrow$

\begin{eqnarray}
\left({\tilde{L}}-{\tilde{S}}{\tilde{E}}\right)^{2}\left({1+\varepsilon}\right)\left({3-\varepsilon}\right) & = & {\tilde{l}}^{2}\label{eq:GKL2SE}\end{eqnarray}
They express the square of the orbital energy and the orbital angular
momentum in terms of the eccentricity, ${\varepsilon}$, and latus
rectum, ${\tilde{l}}$, of a test-particle in its LSO about a KBH
of spin, ${\tilde{S}}$. In equation (\ref{eq:MyL2SE}), the prograde
orbit takes the minus sign and the retrograde orbit takes the plus
sign. The modified form of equation (\ref{eq:MyL2SE}) shown in equation
(\ref{eq:GKL2SE}) corresponds to equation (23) in \cite{2002PhRvD..66d4002G}.

Similar equations derived by Cutler, Kennefick, and Poisson \cite{PhysRevD.50.3816},

\begin{eqnarray}
{\tilde{E}}^{2} & = & \biggl(\Bigl(\tilde{l}-2\left(1+\varepsilon\right)\Bigr)\Bigl(\tilde{l}-2\left(1-\varepsilon\right)\Bigr)\biggr){\tilde{l}}^{-1}{\left(\tilde{l}-3-\varepsilon^{2}\right)}^{-1}\end{eqnarray}
and

\begin{eqnarray}
{\tilde{L}}^{2} & = & {\tilde{l}}^{2}{\left(\tilde{l}-3-\varepsilon^{2}\right)}^{-1},\label{eq:CKPL2}\end{eqnarray}
are only valid for SBH systems. Equation (\ref{eq:MyL2SE}) reduces
to equation (\ref{eq:CKPL2}) when ${\tilde{S}}=0$ and the relationship
${\tilde{l}}=6+2{\varepsilon}$ is used. 

Glampedakis and Kennefick \cite{2002PhRvD..66d4002G} present a similar
treatment which has the advantage of yielding more general results
since it is not assumed that the test-particle has reached the LSO
(i.e. ${r}_{1}>{r}_{3}$). Therefore\begin{eqnarray}
{r}_{3} & = & 2{\left({\tilde{L}}-{\tilde{S}}{\tilde{E}}\right)}^{2}\left(1-{\varepsilon}^{2}\right)\left[{\tilde{l}}^{2}\left(1-{\tilde{E}}^{2}\right)\right]^{-1},\end{eqnarray}
with, ${r}_{1}={R}_{Min}$ and ${r}_{2}={R}_{Max}$, as before. Their
formula for energy,\begin{eqnarray}
{\tilde{E}} & = & \sqrt{1-{\tilde{l}}^{-1}\left(1-{\varepsilon}^{2}\right)\left\{ 1-{\tilde{l}}^{-2}\left({\tilde{L}}-{\tilde{S}}{\tilde{E}}\right)^{2}\left(1-{\varepsilon}^{2}\right)\right\} },\end{eqnarray}
proves to be ideal for generalising our formulae for circular LSOs,
${R}_{LSO}$, to one for elliptical orbits, ${\tilde{l}}_{LSO}$ (See
Appendix C).

\subsubsection{\label{sub:EllipticalLSORadius.}Elliptical LSO Radius.}

The evaluation of ${X}^{2}=\left({\tilde{L}}-{\tilde{S}}{\tilde{E}}\right)^{2}$
in \cite{2002PhRvD..66d4002G} provides a means to extend equations
(\ref{eq:RLSO_Prograde}) and (\ref{eq:RLSO_Retrograde}) beyond their
use with circular LSOs to more general elliptical LSOs by direct substitution
of ${X}^{2}$ into equation (\ref{eq:GKL2SE}). Although a leading
order Taylor expansion (see equation (24) in \cite{2002PhRvD..66d4002G}
) is available from a slow rotation approximation of equation (\ref{eq:GKL2SE})
(i.e. ${\tilde{S}}\thickapprox0$); we present our analytical results.

The general form of the ${\tilde{l}}_{LSO}$ equations for elliptical
orbits are:

\begin{eqnarray}
{\tilde{l}}_{pro} & = & \left(3+\varepsilon\right)+\sqrt{Z_{{o}}}\label{eq:RLSO-Prograde-Elliptical}\\
 & - & \sqrt{16\,{\frac{{\tilde{S}}^{2}\left(1+\varepsilon\right)}{\sqrt{Z_{{o}}}}}-Z_{{o}}+\left(3+\varepsilon\right)^{2}+{\tilde{S}}^{2}\left(1+\varepsilon\right)\left(3-\varepsilon\right)}\nonumber \end{eqnarray}
and\begin{eqnarray}
{\tilde{l}}_{ret} & = & \left(3+\varepsilon\right)+\sqrt{Z_{{o}}}\label{eq:RLSO-Retrograde-Elliptical}\\
 & + & \sqrt{16\,{\frac{{\tilde{S}}^{2}\left(1+\varepsilon\right)}{\sqrt{Z_{{o}}}}}-Z_{{o}}+\left(3+\varepsilon\right)^{2}+{\tilde{S}}^{2}\left(1+\varepsilon\right)\left(3-\varepsilon\right)}\nonumber \end{eqnarray}
Where:

\begin{eqnarray*}
{Z}_{o} & = & 1/3\,{\tilde{S}}^{2}\left(1+\varepsilon\right)\left(3-\varepsilon\right)+1/3\,\left(3+\varepsilon\right)^{2}\\
 & + & 1/3\,{\frac{{\tilde{S}}^{4}\left(1+\varepsilon\right)^{2}\left(3-\varepsilon\right)^{2}-2\,{\tilde{S}}^{2}\left(3+\varepsilon\right)\left(1+\varepsilon\right)\left({\varepsilon}^{2}+15\right)+\left(3+\varepsilon\right)^{4}}{\left(Z_{{i}}\right)^{\left(\frac{1}{3}\right)}}}\\
 & + & 1/3\,\left(Z_{{i}}\right)^{\frac{1}{3}},\end{eqnarray*}

\begin{eqnarray*}
{Z}_{i} & = & \left(3+\varepsilon\right)^{6}\\
 & + & {\tilde{S}}^{2}\left(1+\varepsilon\right)\left({\tilde{S}}^{2}\left(1+\varepsilon\right)\left({\tilde{S}}^{2}\left(1+\varepsilon\right)\left(3-\varepsilon\right)^{3}+3\,{\varepsilon}^{4}+18\,{\varepsilon}^{2}+459\right)-3\,\left({\varepsilon}^{2}+15\right)\left(3+\varepsilon\right)^{3}\right)\\
 & + & 24\,\sqrt{3}\sqrt{Z_{{ii}}},\end{eqnarray*}
and\begin{eqnarray*}
{Z}_{ii} & = & \left(1+\varepsilon\right)^{4}{\tilde{S}}^{6}\left(1-{\tilde{S}}^{2}\right)\left(\left(1-{\varepsilon}\right)\left({\varepsilon}+3\right)^{3}-{\tilde{S}}^{2}\left(1+\varepsilon\right)\left(3-\varepsilon\right)^{3}\right).\end{eqnarray*}
As required, equations (\ref{eq:RLSO-Prograde-Elliptical}) and (\ref{eq:RLSO-Retrograde-Elliptical})
reduce to equations (\ref{eq:RLSO_Prograde}) and (\ref{eq:RLSO_Retrograde})
when ${\varepsilon}=0$. By setting ${\tilde{S}}=0$, both equations
reduce to ${\tilde{l}}=6+2{\varepsilon}$. And in the extreme cases,
where ${\tilde{S}}=1$ (retrograde and prograde), equation (\ref{eq:RLSO-Prograde-Elliptical})
reduces to ${\tilde{l}}=1+{\varepsilon}$; and equation (\ref{eq:RLSO-Retrograde-Elliptical})
reduces to $5+{\varepsilon}+4\sqrt{1+{\varepsilon}}$ , as required.
A detailed treatment of equations (\ref{eq:RLSO-Prograde-Elliptical})
and (\ref{eq:RLSO-Retrograde-Elliptical}) will be outlined in a forthcoming
paper.

\subsection{\label{sub:Calculating-the-LSO}Calculating the LSO Properties}

\subsubsection{Introduction.}

The elements have now been found to perform general calculations of
the LSO for arbitrary values of KBH spin, ${\tilde{S}}$, orbital
angular momentum, ${\tilde{L}}$, and total energy, ${\tilde{E}}$.
Although we have analytical formulae that give us ${\tilde{l}}_{LSO}$
for general elliptical orbits, it is important to construct and outline
our methodology in preparation for future work on test-particle orbits
that are inclined with respect to the equatorial plane of the KBH.
We must quantify the relationship between the value of ${\tilde{L}}$
for the test-particle orbit and the shape of its effective potential
surface. 

Here we outline, in detail, our numerical method for calculating the
latus rectum, ${\tilde{l}}$, and eccentricity, ${\varepsilon}$,
of LSO orbits. These values will help us to appraise the usefulness
of our new, generalised ${\tilde{l}}_{LSO}$ equations in (\ref{eq:RLSO-Prograde-Elliptical})
and (\ref{eq:RLSO-Retrograde-Elliptical}).

\subsubsection{\label{sub:Algorithm}Algorithm.}

For clarity, an example where ${\tilde{S}}=0.5$ and the test-particle
is in a prograde orbit is demonstrated. In Table \ref{tab:LSO-Properties},
the calculations for a retrograde orbit, and an SBH are included for
comparison. 

Such an algorithm proceeds as follows:

\paragraph{Specify the KBH spin -}

A given problem will most likely have a prior specification of a fixed
value of ${\tilde{S}}$, where $0\leq{\tilde{S}}<1$ for either a
prograde or retrograde orbit (if a retrograde orbit is used, \textit{(ret),}
will follow the value assigned to ${\tilde{S}}$). In this example
we shall use ${\tilde{S}}=0.5$ and a prograde orbit since it has
already been used in the calculation of ${R}_{LSO}$ and ${\tilde{L}}$
for prograde and retrograde LSOs previously in this paper (see \ref{sub:LastStableOrbit}
and Figure \ref{fig:RvsL}). Similar calculations were performed for
the ${\tilde{S}}=0.5(ret)$ case, and for an SBH (See Table \ref{tab:LSO-Properties}). 

We use either equation (\ref{eq:RLSO_Prograde}) for a prograde orbit
or equation (\ref{eq:RLSO_Retrograde}) for a retrograde orbit to
directly calculate ${R}_{LSO}$ (BL coordinates) thus,

\begin{eqnarray}
{\tilde{S}}=0.5 & \Rightarrow & {R}_{LSO}=4.23300,\end{eqnarray}
which gives us the LSO radius of a circular orbit, ${R}_{LSO}$.

\paragraph{Find $\tilde{L}$ -}

The values of ${\tilde{S}}$ and ${R}_{LSO}$ can now be used to calculate
the value of ${\tilde{L}}^{2}$ assuming the LSO is at a point of
inflection (i.e. a circular orbit) \textit{vis.} equations (\ref{eq:L2-Quadratic-Solution-Pro})
or (\ref{eq:L2-Quadratic-Solution-Retro}) depending on the direction
of the orbit. The result for ${\tilde{S}}=0.5$ and ${R}_{LSO}=4.23300$
is found to be,

\begin{eqnarray}
{\tilde{L}}^{2} & = & 8.4266319.\end{eqnarray}
This value is necessarily a positive quantity, hence the need to ensure
that the correct prograde or retrograde orbital angular momentum equation
has been used.

\paragraph{Calculate ${\tilde{E}}$ -}

Because the values of ${\tilde{S}},{R}_{LSO},$ and ${\tilde{L}}$
are known at the point of inflection, we can use ${\tilde{V}}_{+}$
(see equation (\ref{eq:PotentialPlusMinus})) to directly calculate
the energy, ${\tilde{E}}$, of the test-particle in a circular orbit,
i.e.

\begin{eqnarray}
{\tilde{E}} & = & 0.91788201.\end{eqnarray}
N.B., the value of ${\tilde{E}}<{1.0}$, hence the orbit is bound.
Whenever ${\tilde{E}}\geqq{1.0}$, the orbit is not bound.

\paragraph{Expand to include elliptical orbits -}

By careful examination of Figure \ref{fig:Min-of-L-wrt-R} one sees
that the local minimum of ${\tilde{L}}$ corresponds to the case where
the LSO is circular; the values of ${R}_{LSO}$ and the radius of
the local minimum of the potential, ${\tilde{V}}_{+}$, (Figure \ref{fig:Potentials_For_Different_L})
coincide, as expected. The angular momentum, ${\tilde{L}},$ that
corresponds to an elliptical LSO is then higher than that for a circular
LSO.

The algorithm shall be broadened to include the case of an elliptical
orbit. For the orbit to be elliptical, the orbital angular momentum,
(${\tilde{L}}=\sqrt{{\tilde{L}}^{2}}\Rightarrow{\tilde{L}}=\sqrt{8.4266319}$)
must be increased by an arbitrary factor ${\delta\tilde{L}}$ (where
${\delta\tilde{L}}>0$, see Figure \ref{fig:Min-of-L-wrt-R}); the
slight increase in ${\tilde{L}}$ above its minimum value changes
the LSO from a circular orbit, to one that is elliptical. Accordingly
the value of ${\tilde{E}}$ will increase and the value of ${R}_{Min}$
will be reduced. A similar treatment of elliptical orbits, based upon
increments of ${\tilde{L}}$, can be found in \cite{2000PhRvD..62l4022O}.

\paragraph{Find ${R}_{Min}$ for the elliptical orbit -}

By working with a larger value of orbital angular momentum in the
form (${\tilde{L}}_{Elliptical}={\tilde{L}}_{Circular}+{\delta\tilde{L}}$)
we can calculate the value of ${R}_{Min}$ without requiring the new
value of the orbital energy, ${\tilde{E}}$ (see Figure \ref{fig:Min-of-L-wrt-R}).
If ${\delta\tilde{L}}=0.01$, then ${\tilde{L}}=2.904588078$; therefore,
(\textit{vis.} equations (\ref{eq:L2-Quadratic-Solution-Pro}) or
(\ref{eq:L2-Quadratic-Solution-Retro})) the new value of ${R}_{Min}$
can be calculated numerically to yield:

\begin{equation}
{R}_{Min}=4.10329200.\end{equation}
Correspondingly, the total orbital energy can be calculated (\textit{vis.}
equation (\ref{eq:PotentialPlusMinus})):

\begin{eqnarray}
{\tilde{E}}_{LSO}^{Elliptical} & = & 0.9180746,\end{eqnarray}
\textit{cf.,}\begin{eqnarray}
{\tilde{E}}_{LSO}^{Circular} & = & 0.91788201.\end{eqnarray}
As required: ${\tilde{E}}_{LSO}^{Elliptical}>{\tilde{E}}_{LSO}^{Circular}$.

\paragraph{Find the maximum radius for an elliptical orbit -}

In calculating a data set, the various values of ${\delta\tilde{L}}$
are selected and the corresponding values of ${\varepsilon}$ and
${\tilde{l}}$ are found. Now that the value of ${\tilde{E}}_{LSO}^{Elliptical}$
is known, the maximum radius of the elliptical orbit $\left({R}_{Max}\right)$
can be calculated numerically, \textit{vis.} ${\tilde{V}}_{+}={\tilde{E}}_{LSO}^{Elliptical}$,
because the effective potential of the test-particle has the same
value at ${R}_{Min}$ and ${R}_{Max}$. The result is:\begin{equation}
{R}_{Max}=4.520999771.\end{equation}

\paragraph{Determine the elliptical orbit parameters -}

We now have the information necessary to calculate the latus rectum,
${\tilde{l}}$, and the eccentricity, $\varepsilon$, of the orbit.
The dimensionless semi-major axis, ${A}$, of the elliptical orbit
can be calculated from the values of ${R}_{LSO}$ and ${R}_{Max}$:\begin{eqnarray}
A & = & \frac{\left({R}_{Min}+{R}_{Max}\right)}{2},\nonumber \\
 & = & 4.3121.\end{eqnarray}
 \begin{eqnarray}
\varepsilon & = & 1-\frac{{R}_{Min}}{{A}},\nonumber \\
 & = & 0.0484.\label{eq:epsilonformula}\end{eqnarray}
The latus rectum, ${\tilde{l}}$, is now be calculated,

\begin{eqnarray}
{\tilde{l}} & = & {A}\left(1-{\varepsilon}^{2}\right),\nonumber \\
 & = & 4.3020.\label{eq:LatusRectumFormula}\end{eqnarray}
N.B., the values of ${A}$, ${\tilde{l}}$, ${R}_{LSO}$ (for circular
LSO), ${R}_{Min}$ and ${R}_{Max}$ are expressed in terms of BL coordinates.

\subsubsection{Calculation of the normalised orbital frequency}

According to the relativistic form of Kepler's third law (see problem
17.4 in \cite{1975STIA...7626675L} or exercise 12.7 in \cite{1983bhwd.book.....S}),
the orbital period of a closed orbit, ${P}$, can be expressed in
terms of the semi-major axis of the orbit, ${a}$, and the mass, ${M}$
of the central body about which the test-particle orbits. This equation
applies to elliptical orbits in general. If the orbit is subject to
precession, then the value, ${\nu}$, represents the orbital frequency
of the test-particle in an open orbit. Hence,

\begin{eqnarray}
P & = & 2{\pi}\,\left|{\frac{{a}^{3/2}\pm\left|{\mathbf{s}}\right|{\sqrt{M}}}{\sqrt{M}}}\right|,\end{eqnarray}
where ${\mathbf{s}}={\mathbf{J}}/{M}$; and the plus sign corresponds
to the prograde orbit and the minus sign corresponds to the retrograde
orbit . The corresponding orbital frequency is:

\begin{eqnarray}
{\nu} & = & {P}^{-1};\end{eqnarray}
therefore,\begin{eqnarray}
{\nu} & = & \frac{1}{2\pi}\,\left|{\frac{\sqrt{M}}{\left({a}^{3/2}\pm\left|{\mathbf{s}}\right|{\sqrt{M}}\right)}}\right|,\end{eqnarray}
which leads to,\begin{eqnarray}
W & = & 2\pi{M}{\nu}\nonumber \\
 & = & \left|{\frac{{M}^{3/2}}{\left({a}^{3/2}\pm\left|{\mathbf{s}}\right|{\sqrt{M}}\right)}}\right|,\label{eq:Keplers-W-a}\end{eqnarray}
(in equation (\ref{eq:Keplers-W-a}) the parameter, ${a}$, refers
to the length of the semi-major axis). When variables normalised with
respect to the KBH mass, ${M}$, are used:

\begin{eqnarray}
{A} & = & \frac{{a}}{{M}},\end{eqnarray}
and\begin{eqnarray}
{\tilde{S}} & = & \left|{\mathbf{s}}\right|/M;\end{eqnarray}
one obtains,\begin{eqnarray}
W & =\left|\frac{1}{{A}^{3/2}\pm{\tilde{S}}}\right| & .\label{eq:WFormulaA}\end{eqnarray}
If equation (\ref{eq:LatusRectumFormula}) is then used to represent
equation (\ref{eq:WFormulaA}) in terms of the dimensionless latus
rectum, ${\tilde{l}}$:

\begin{eqnarray}
{W} & =\left|{\left(1-{\varepsilon}^{2}\right)}^{\frac{3}{2}}/\left({\tilde{l}}^{\frac{3}{2}}\pm{\tilde{S}}{\left(1-{\varepsilon}^{2}\right)}^{\frac{3}{2}}\right)\right| & .\label{eq:WFormulaL}\end{eqnarray}
\begin{sidewaystable}
\caption{\label{tab:LSO-Properties}LSO parameters calculated for both circular
and elliptical orbits where ${\tilde{S}}=0.5\left(ret\right),\:0.0,$
and $0.5$ to four decimal places; these values may be carried to
greater precision. }

\begin{tabular}{cccccccc}
 & \multicolumn{3}{c}{Circular Orbit} &  & \multicolumn{3}{c}{Elliptical Orbit}\tabularnewline
\hline 
KBH Spin (${\tilde{S}}$) & $0.5(ret)$ & $0.0$ & $+0.5$ &  & $0.5(ret)$ & $0.0$ & $+0.5$\tabularnewline
\hline
\cline{1-1} \cline{2-4} \cline{5-5} 
${\delta\tilde{L}}$ & \multicolumn{3}{c}{$0.00$} &  & \multicolumn{3}{c}{$0.01$}\tabularnewline
\cline{1-1} \cline{2-4} \cline{5-5} 
\hline 
$R_{LSO}$ & $7.5546$ & $6.0000$ & $4.2330$ & ${R}_{Min}$ & $7.3576$ & $5.8317$ & $4.1033$\tabularnewline
${\tilde{L}}^{2}$ & $15.0871$ & $12.0000$ & $8.4266$ &  & $15.0971$ & $12.0100$ & $8.4266$\tabularnewline
${\tilde{E}}_{LSO}$ & $0.9728$ & $0.9428$ & $0.9179$ &  & $0.9549$ & $0.9429$ & $0.9181$\tabularnewline
$R_{Max}$ & $7.5546$ & $6.0000$ & $4.23300$ &  & $7.9806$ & $6.3675$ & $4.5210$\tabularnewline
${A}_{BL}$ & $7.5546$ & $6.0000$ & $4.23300$ &  & $7.66912$ & $6.0996$ & $4.31215$\tabularnewline
${\varepsilon}_{BL}$ & $0.0000$ & $0.0000$ & $0.0000$ &  & $0.04062$ & $0.04392$ & $0.0484$\tabularnewline
${\tilde{l}}_{BL}$ & $7.5546$ & $6.0000$ & $4.2330$ &  & $7.6565$ & $6.0880$ & $4.3020$\tabularnewline
${W}_{BL}$ & $0.04935$ & $0.06804$ & $0.10856$ &  & $0.04822$ & $0.06638$ & $0.10577$\tabularnewline
${A}_{spherical}$ & $7.57111$ & $6.0000$ & $4.2624$ &  & $7.68512$ & $6.09960$ & $4.34110$\tabularnewline
${\varepsilon}_{spherical}$ & $0.0000$ & $0.0000$ & $0.0000$ &  & $0.04045$ & $0.0439$ & $0.0478$\tabularnewline
${\tilde{l}}_{spherical}$ & $7.57111$ & $6.0000$ & $4.26243$ &  & $7.67284$ & $6.087834$ & $4.3312$$ $\tabularnewline
${W}_{spherical}$ & $0.04919$ & $0.06804$ & $0.10753$ &  & $0.04806$ & $0.06638$ & $0.10477$\tabularnewline
\hline
\end{tabular}
\end{sidewaystable}

\subsection{\label{sub:Calculations}Calculations}

\subsubsection{LSO and orbit characteristics.}

Three methods were used to calculate ${\tilde{l}}_{LSO}$ for orbits
of various eccentricity ($0\leq\varepsilon\leq1.0$) and KBH spin
(${\tilde{S}}=0.5,0.99$; prograde and retrograde). The values obtained
here are shown alongside those found in the literature \cite{2000PhRvD..62l4022O,2002PhRvD..66d4002G}
in Tables \ref{tab:RetrogradeData} and \ref{tab:ProgradeData}. The
${\tilde{l}}_{LSO}$ values we obtained by following the algorithm
described in \ref{sub:Algorithm} are listed in the Numerical column.
The general formulae described in \ref{sub:EllipticalLSORadius.}
were used to generate the values in the Analytical column. A third
method was used to numerically estimate the ${\tilde{l}}_{LSO}$ values
directly from the companion matrix (Appendix C) by first substituting
the ${\tilde{S}}$ and ${\varepsilon}$ values into the matrix before
calculating its eigenvalues.

\begin{sidewaystable}
\caption{\label{tab:RetrogradeData}Retrograde LSO data from \cite{2000PhRvD..62l4022O,2002PhRvD..66d4002G}
and results calculated using the numerical and analytical methods
presented in this work (our results were rounded off to four decimal
places).}
\begin{tabular}{ccccccccccc}
\toprule 
 & Ret Limit & \multicolumn{4}{c}{${\tilde{l}}$ for $\tilde{S}=0.99(ret)$} & \multicolumn{4}{c}{${\tilde{l}}$ for $\tilde{S}=0.5(ret)$} & ${\tilde{S}}=0$\tabularnewline
\midrule
${\varepsilon}$ & \begin{tabular}{c}
${\tilde{l}}=5+\varepsilon$\tabularnewline
$+4\sqrt{1+\varepsilon}$\tabularnewline
\end{tabular} & \begin{tabular}{c}
\cite{2002PhRvD..66d4002G}\tabularnewline
$^{\star}$\cite{2000PhRvD..62l4022O}\tabularnewline
\end{tabular} & {\small Numerical} & {\small Analytical} & {\small }\begin{tabular}{c}
{\small Companion}\tabularnewline
{\small Matrix}\tabularnewline
\end{tabular} & \begin{tabular}{c}
\tabularnewline
$^{\star}$\cite{2000PhRvD..62l4022O}\tabularnewline
\end{tabular} & {\small Numerical} & {\small Analytical} & {\small }\begin{tabular}{c}
{\small Companion}\tabularnewline
{\small Matrix}\tabularnewline
\end{tabular} & ${\tilde{l}}=6+2{\varepsilon}$\tabularnewline
\midrule
$0.0$ & $9.0000$ & $8.972^{\star}$  & $8.9719$ & $8.9719$ & $8.9719$ & $7.555^{\star}$  & $7.5549$ & $7.5546$ & $7.5546$ & $6.00$\tabularnewline
$0.1$ & $9.2952$ & $9.266$ & $9.2662$ & $9.2662$ & $9.2662$ & - & $7.8040$ & $7.8039$ & $7.8039$ & $6.20$\tabularnewline
$0.2$ & $9.5818$ & $9.552$ & $9.5519$ & $9.5519$ & $9.5519$ & - & $8.0489$ & $8.0488$ & $8.0488$ & $6.40$\tabularnewline
$0.3$ & $9.8607$ & $9.83$ & $9.8301$ & $9.8301$ & $9.8301$ & - & $8.2922$ & $8.2899$ & $8.2899$ & $6.60$\tabularnewline
$0.4$ & $10.1329$ & $10.102$ & $10.1017$ & $10.1016$ & $10.1016$ & - & $8.5275$ & $8.5274$ & $8.5275$ & $6.80$\tabularnewline
$0.5$ & $10.3990$ & $10.367$ & $10.3671$ & $10.3671$ & $10.3671$ & - & $8.7620$ & $8.7620$ & $8.7620$ & $7.00$\tabularnewline
$0.6$ & $10.6596$ & $10.627$ & $10.6272$ & $10.6272$ & $10.6272$ & - & $8.9940$ & $8.9939$ & $8.9939$ & $7.20$\tabularnewline
$0.7$ & $10.9154$ & $10.882$ & $10.8824$ & $10.8824$ & $10.8824$ & - & $9.2233$ & $9.2233$ & $9.2233$ & $7.40$\tabularnewline
$0.8$ & $11.1666$ & $11.133$ & $11.1332$ & $11.1332$ & $11.1332$ & - & $9.4506$ & $9.4505$ & $9.4505$ & $7.60$\tabularnewline
$0.9$ & $11.4136$ & $11.38$ & $11.3798$ & $11.3798$ & $11.3798$ & - & $9.6756$ & $9.6756$ & $9.6756$ & $7.80$\tabularnewline
$1.0$ & $11.6569$ & $11.623$ & $11.6226$ & $11.6227$ & $11.6227$ & - & $9.8988$ & $9.8990$ & $9.8990$ & $8.00$\tabularnewline
\bottomrule 
\end{tabular}
\end{sidewaystable}

\begin{sidewaystable}
\caption{\label{tab:ProgradeData}Prograde LSO data from \cite{2000PhRvD..62l4022O,2002PhRvD..66d4002G}
and results calculated using the numerical and analytical methods
presented in this work (our results were rounded off to four decimal
places).}
\begin{tabular}{ccccccccccc}
\toprule 
 & ${\tilde{S}}=0$ & \multicolumn{4}{c}{${\tilde{l}}$ for $\tilde{S}=0.5$} & \multicolumn{4}{c}{${\tilde{l}}$ for$\tilde{S}=0.99$} & Pro Limit\tabularnewline
\midrule 
${\varepsilon}$ & ${\tilde{l}}=6+2{\varepsilon}$ & \begin{tabular}{c}
\cite{2002PhRvD..66d4002G}\tabularnewline
$^{\star}$\cite{2000PhRvD..62l4022O}\tabularnewline
\end{tabular} & {\small Numerical} & {\small Analytical} & {\small }\begin{tabular}{c}
{\small Companion}\tabularnewline
{\small Matrix}\tabularnewline
\end{tabular} & \begin{tabular}{c}
\cite{2002PhRvD..66d4002G}\tabularnewline
$^{\star}$\cite{2000PhRvD..62l4022O}\tabularnewline
\end{tabular} & {\small Numerical} & {\small Analytical} & {\small }\begin{tabular}{c}
{\small Companion}\tabularnewline
{\small Matrix}\tabularnewline
\end{tabular} & ${\tilde{l}}=1+{\varepsilon}$\tabularnewline
\midrule
$0.0$ & $6.00$ & $4.233^{\star}$  & $4.2330$ & $4.2330$ & $4.2330$ & $1.454^{\star}$ & $1.4545$ & $1.4545$ & $1.4545$ & $1.00$\tabularnewline
$0.1$ & $6.20$ & $4.377$ & $4.3769$ & $4.3769$ & $4.3769$ & $1.516$ & $1.5156$ & $1.5156$ & $1.5156$ & $1.10$\tabularnewline
$0.2$ & $6.40$ & $4.526$ & $4.5259$ & $4.5259$ & $4.5259$ & $1.595$ & $1.5950$ & $1.5950$ & $1.5950$ & $1.20$\tabularnewline
$0.3$ & $6.60$ & $4.679$ & $4.6792$ & $4.6792$ & $4.6792$ & $1.685$ & $1.6852$ & $1.6852$ & $1.6852$ & $1.30$\tabularnewline
$0.4$ & $6.80$ & $4.836$ & $4.8360$ & $4.8360$ & $4.8360$ & $1.782$ & $1.7822$ & $1.7822$ & $1.7822$ & $1.40$\tabularnewline
$0.5$ & $7.00$ & $4.996$ & $4.9959$ & $4.9959$ & $4.9959$ & $1.883$ & $1.8835$ & $1.8835$ & $1.8835$ & $1.50$\tabularnewline
$0.6$ & $7.20$ & $5.158$ & $5.1584$ & $5.1584$ & $5.1584$ & $1.988$ & $1.9876$ & $1.9876$ & $1.9876$ & $1.60$\tabularnewline
$0.7$ & $7.40$ & $5.323$ & $5.3232$ & $5.3233$ & $5.3232$ & $2.094$ & $2.0954$ & $2.0939$ & $2.0939$ & $1.70$\tabularnewline
$0.8$ & $7.60$ & $5.49$ & $5.4899$ & $5.4901$ & $5.4900$ & $2.201$ & $2.2016$ & $2.2016$ & $2.2016$ & $1.80$\tabularnewline
$0.9$ & $7.80$ & $5.658$ & $5.6584$ & $5.6586$ & $5.6584$ & $2.31$ & $2.3104$ & $2.3104$ & $2.3104$ & $1.90$\tabularnewline
$1.0$ & $8.00$ & $5.828$ & $5.8284$ & $5.8287$ & $5.8284$ & $2.42$ & $2.4199$ & $2.4200$ & $2.4200$ & $2.00$\tabularnewline
\bottomrule 
\end{tabular}
\end{sidewaystable}

The agreement between our various calculation methods, and with the
results published previously in \cite{2000PhRvD..62l4022O,2002PhRvD..66d4002G}
is excellent (i.e. error $<0.1\%$). Therefore the algorithmic method
we have outlined in \ref{sub:Algorithm} may be considered reliable.
And the use of the companion matrix (see Appendix B and C) in performing
numerical calculations of the LSO parameters has been successfully
demonstrated.

\subsubsection{\label{sub:Conversion-BL-Spherical}Conversion from the BL to the
spherical coordinate system.}

The foregoing analysis was performed in the BL coordinate system in
which we suppressed the use of the, BL, subscript. To apply these
estimates of the LSO parameters in the problem of setting the boundary
conditions needed in modelling the evolution equations, it is necessary
to convert them to the spherical coordinate system. We shall describe
this conversion process, and state the appropriate caveats. 

Equation (\ref{eq:RSpherical-Equatorial}) provides the means to convert
any radial distance on an elliptical orbit (that lies in the equatorial
plane of the KBH) expressed in BL coordinates into a radial distance
in spherical (or cylindrical) coordinates. But one cannot proceed
precipitously; an elliptical orbit in the BL coordinate system, will
be only a good approximation of an ellipse once expressed in the spherical
coordinate system. In addition, careful consideration must be given
to the values of ${\phi}$ in their respective coordinate systems
as there will be some important differences that will demand a more
profound understanding and a more cautious interpretation.

Consider the case of a test-particle in an elliptical orbit about
a KBH. The absence of the parameter, ${\phi}$, from equation (\ref{eq:RSpherical-Equatorial})
notwithstanding; the angle, 

\begin{eqnarray}
{\phi}_{spherical} & = & {\phi}_{BL}\mp\arctan\left(\frac{{\tilde{S}}}{{R}_{BL}}\right)\label{eq:PhiBLtoSph}\end{eqnarray}
\textit{vis.} equations (\ref{eq:BL_coordinates}) and (\ref{eq:arctanSR}),
will force the points on the orbit that correspond to ${R}_{LSO\,\left(spherical\right)}$,
${R}_{Max\,\left(spherical\right)}$, and the position of the MBH
at the focus of the ellipse, to be no longer collinear. Therefore
the use of the values of ${R}_{LSO\,\left(spherical\right)}$ and
${R}_{Max\,\left(spherical\right)}$ to calculate ${\varepsilon}_{spherical}$
(\textit{vis.} equation (\ref{eq:epsilonformula})) is potentially
a source of error, especially for KBHs of large spin.

We calculate:

\begin{eqnarray}
{R}_{Min\,\left(spherical\right)} & = & \sqrt{{R}_{Min\,\left(BL\right)}^{2}+{\tilde{S}}^{2}}\end{eqnarray}
and

\begin{eqnarray}
{R}_{Max\,\left(spherical\right)} & = & \sqrt{{R}_{Max\,\left(BL\right)}^{2}+{\tilde{S}}^{2}}.\end{eqnarray}
These two values are used to calculate the semi-major axis:

\begin{eqnarray}
{A}_{Spherical} & = & \frac{{R}_{Min\,\left(spherical\right)}+{R}_{Max\,\left(spherical\right)}}{2},\end{eqnarray}
from which one can obtain\begin{eqnarray}
{\varepsilon}_{spherical} & = & 1-\frac{{R}_{Min\,\left(spherical\right)}}{{A}_{LSO\,\left(spherical\right)}}.\end{eqnarray}
The latus rectum can be calculated using,\begin{eqnarray}
{\tilde{l}}_{spherical} & = & {A}_{spherical}\left(1-{\varepsilon}_{spherical}^{2}\right),\end{eqnarray}
 which is analogous to equation (\ref{eq:LatusRectumFormula}). The
orbital frequency, ${W}_{spherical}$, is obtained from equation (\ref{eq:WFormulaA}).
The values of these parameters expressed in spherical coordinates
are reported in Table \ref{tab:LSO-Properties}. 

\begin{figure}
\caption{\label{fig:Comparison-of-Orbits0.50}A comparison of orbits in BL
and spherical coordinates for a KBH of spin, ${\tilde{S}}=0.5$ (prograde
and retrograde). The view is taken from above the KBH equatorial plane.
Orbital precession is not included.}

\subfigure[Prograde]{\includegraphics[width=0.6\paperwidth]{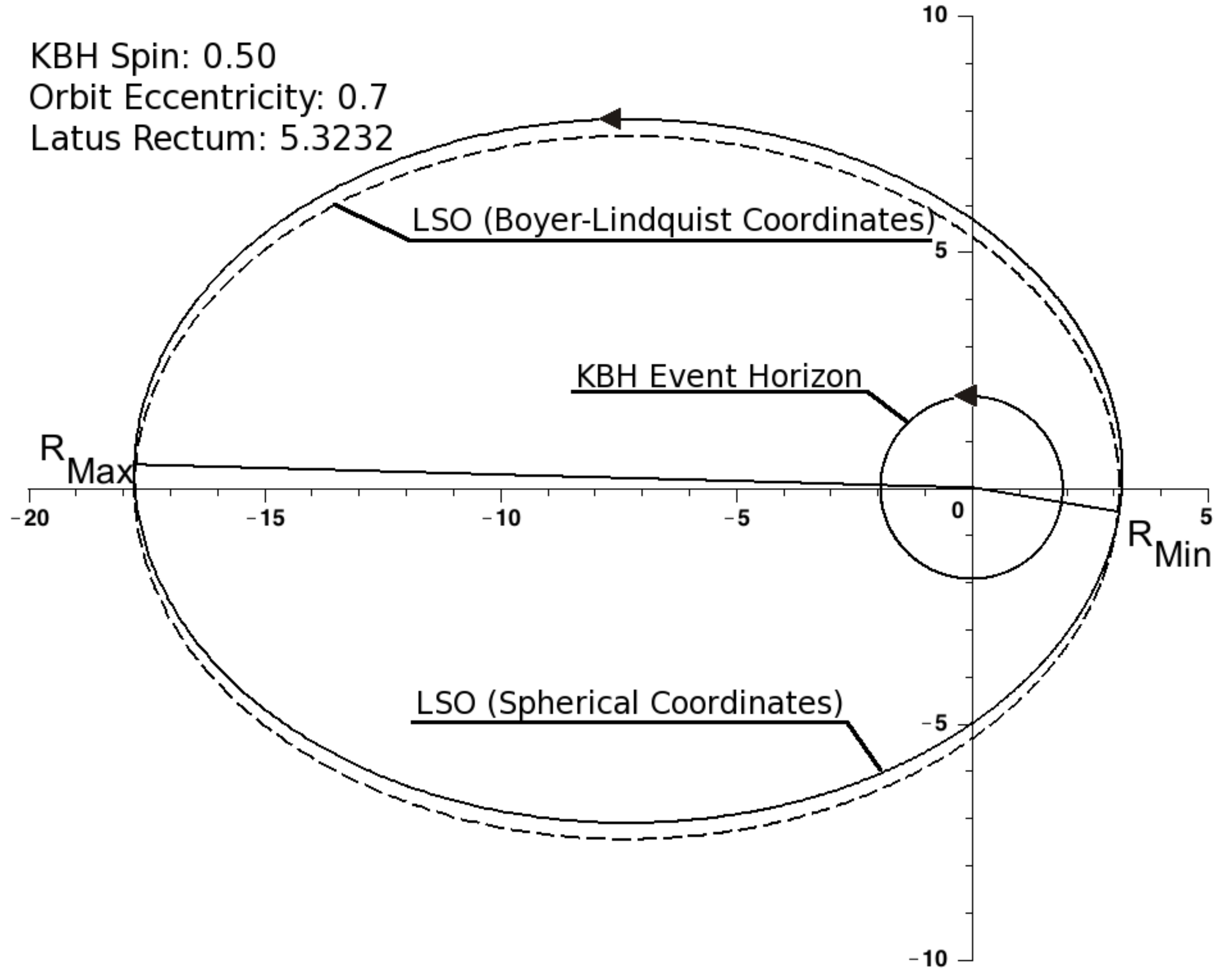}}

\subfigure[Retrograde]{\includegraphics[width=0.6\paperwidth]{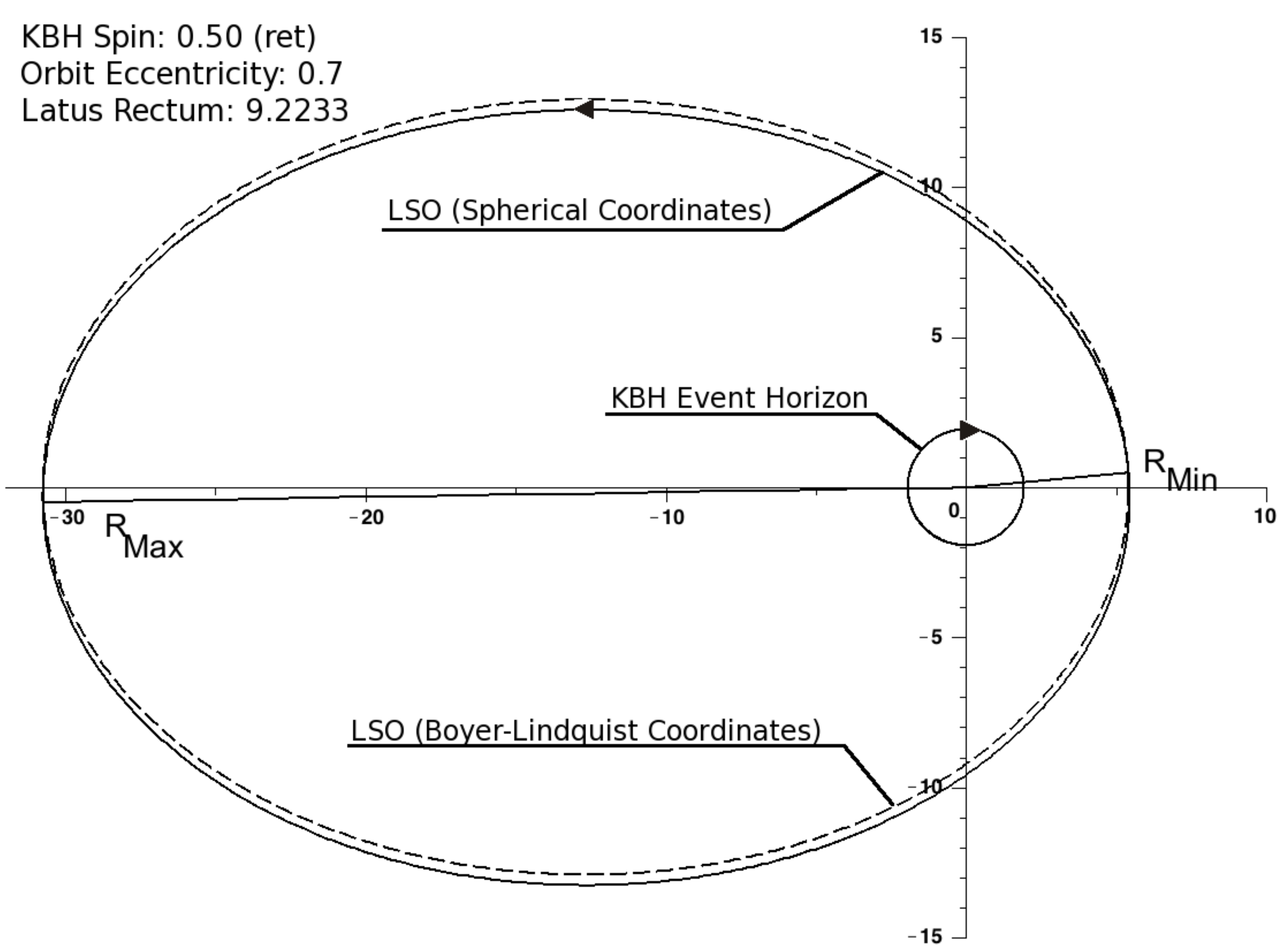}}

\end{figure}

\begin{figure}
\caption{\label{fig:Comparison-of-Orbits0.99}A comparison of orbits in BL
and spherical coordinates for a KBH of spin, ${\tilde{S}}=0.99$ (prograde
and retrograde). The view is taken from above the KBH equatorial plane.
Orbital precession is not included.}

\subfigure[Prograde]{\includegraphics[width=0.6\paperwidth]{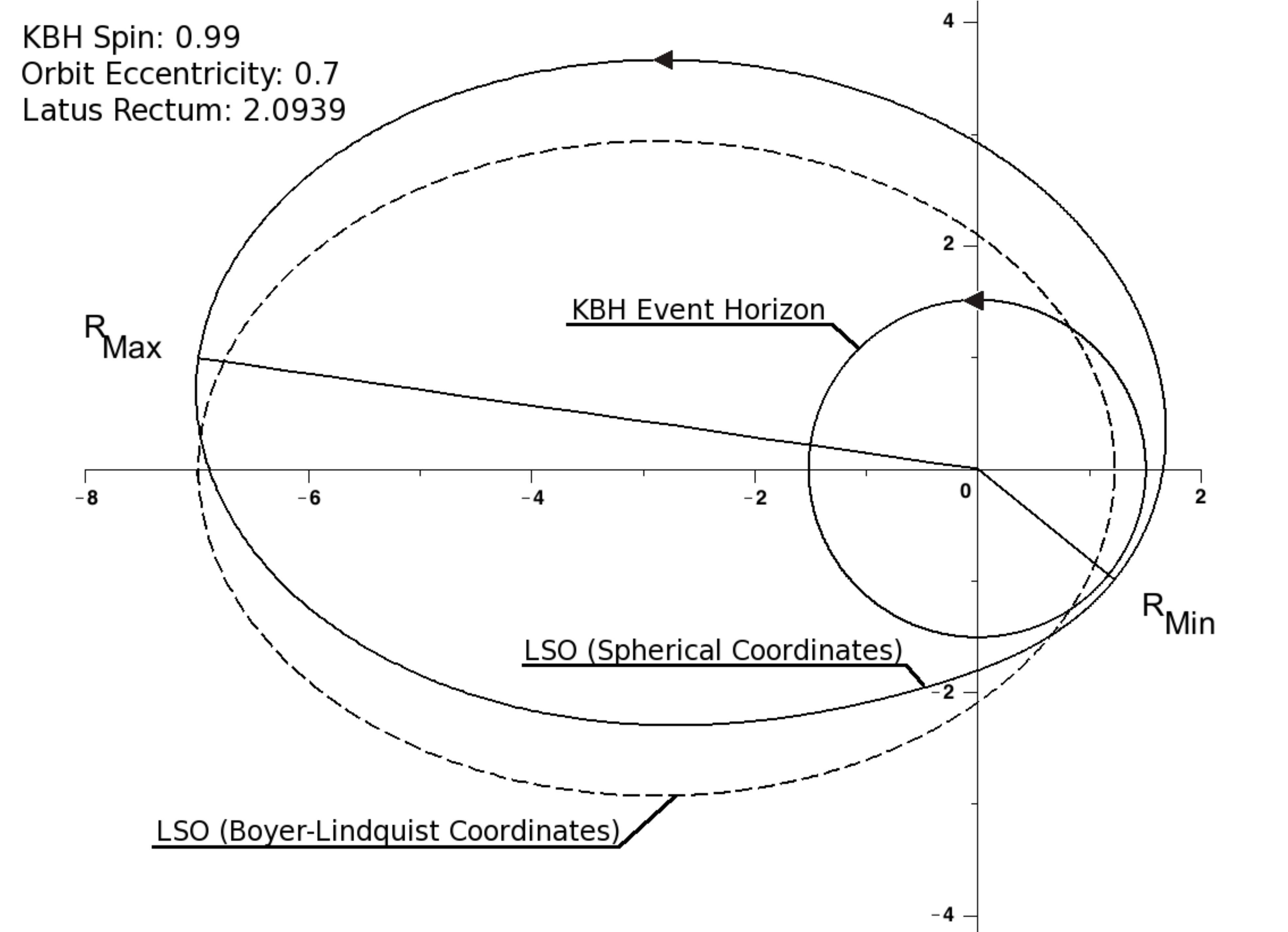}}

\subfigure[Retrograde]{\includegraphics[width=0.6\paperwidth]{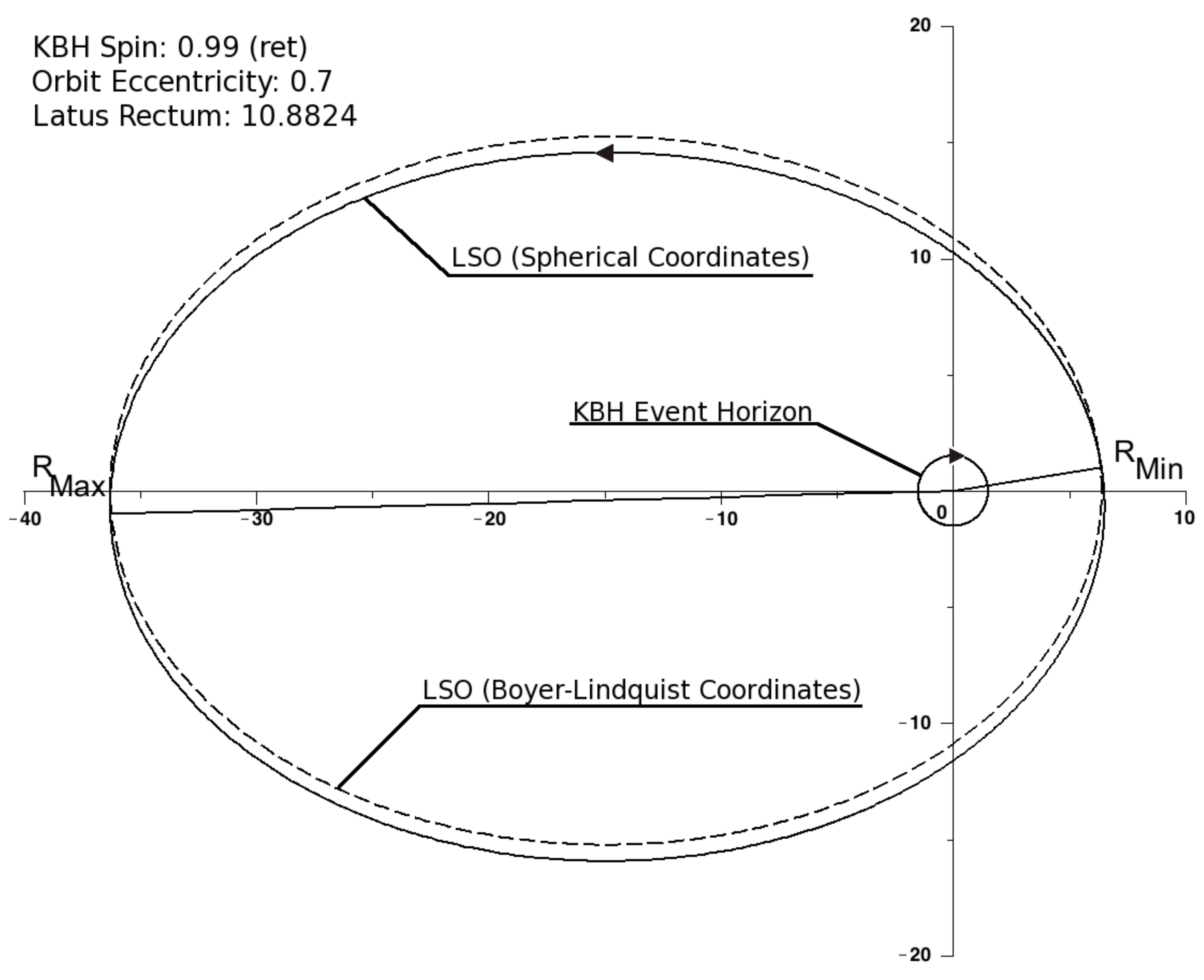}}

\end{figure}

The behaviour of ${\phi}_{spherical}$ is not part of this study;
but further investigation will be undertaken since an understanding
of ${\phi}_{spherical}$ is essential for properly characterising
the zoom and whirl of the test-particle in its orbit. A diagrammatic
comparison of test-particle orbits in the BL and spherical coordinate
systems is shown in Figures \ref{fig:Comparison-of-Orbits0.50} and
\ref{fig:Comparison-of-Orbits0.99} for a KBH of spins of ${\tilde{S}}=0.5$
and ${\tilde{S}}=0.99$ respectively. The orbit parameters are taken
from Tables \ref{tab:RetrogradeData} and \ref{tab:ProgradeData}
for $\varepsilon=0.7$. One can view the shift in the value of ${\phi}_{spherical}$
as arising from the Lense-Thirring precession \cite{1997PhLA..233...25R};
the orbit has a shape that can be approximated as an ellipse that
is precessing. The orbit diagrams shown in Figures \ref{fig:Comparison-of-Orbits0.50}
and \ref{fig:Comparison-of-Orbits0.99} exclude this orbital precession.

\subsubsection{LSO formulae.}

\begin{sidewaysfigure}
\includegraphics[width=0.8\textwidth]{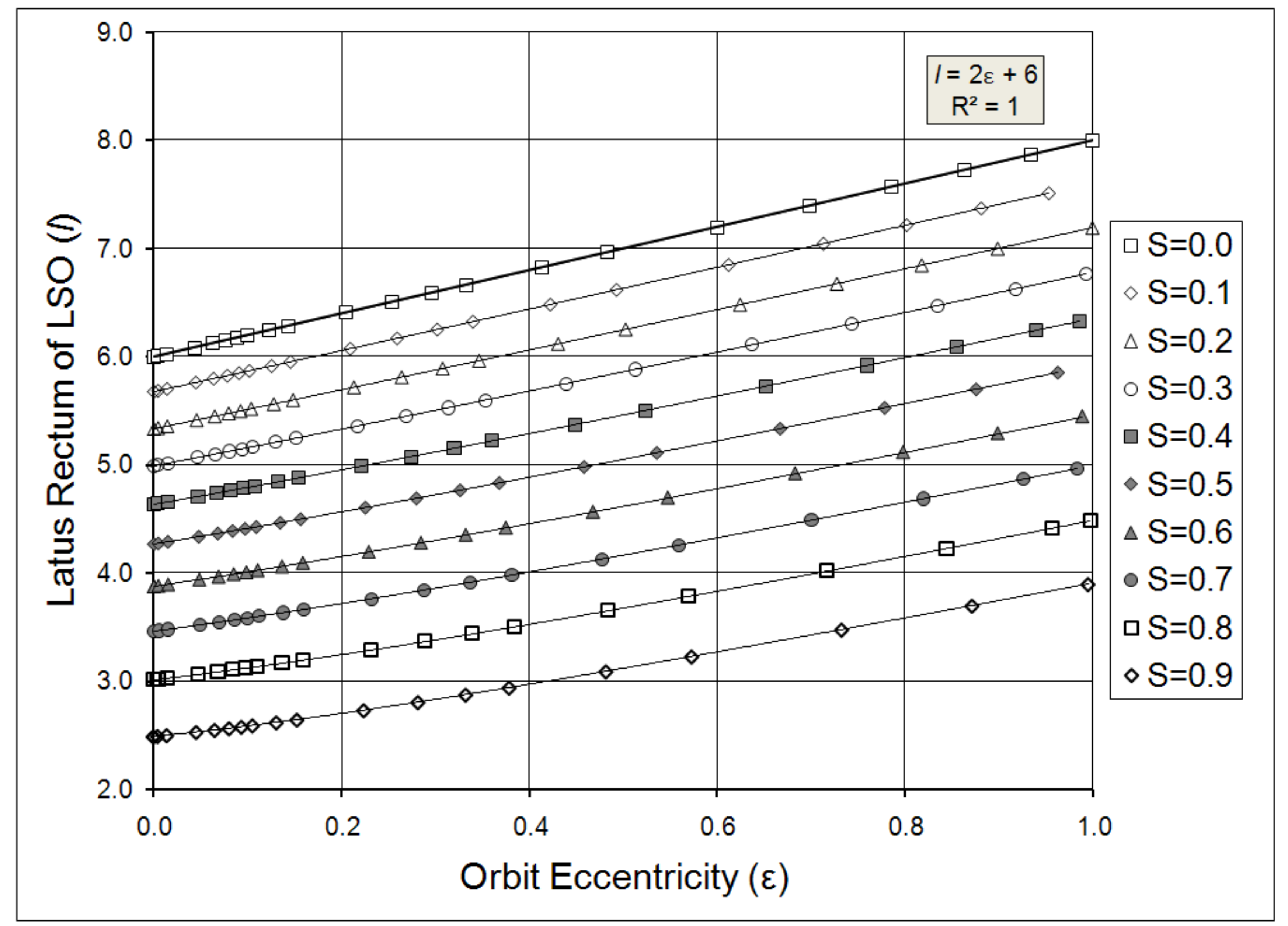}\caption{\label{fig:Prograde-LSO}The LSO latus rectum, ${\tilde{l}}$, calculated
for KBH systems in which the test-particle is in a prograde orbit.}

\end{sidewaysfigure}

\begin{sidewaysfigure}
\includegraphics[width=0.8\textwidth]{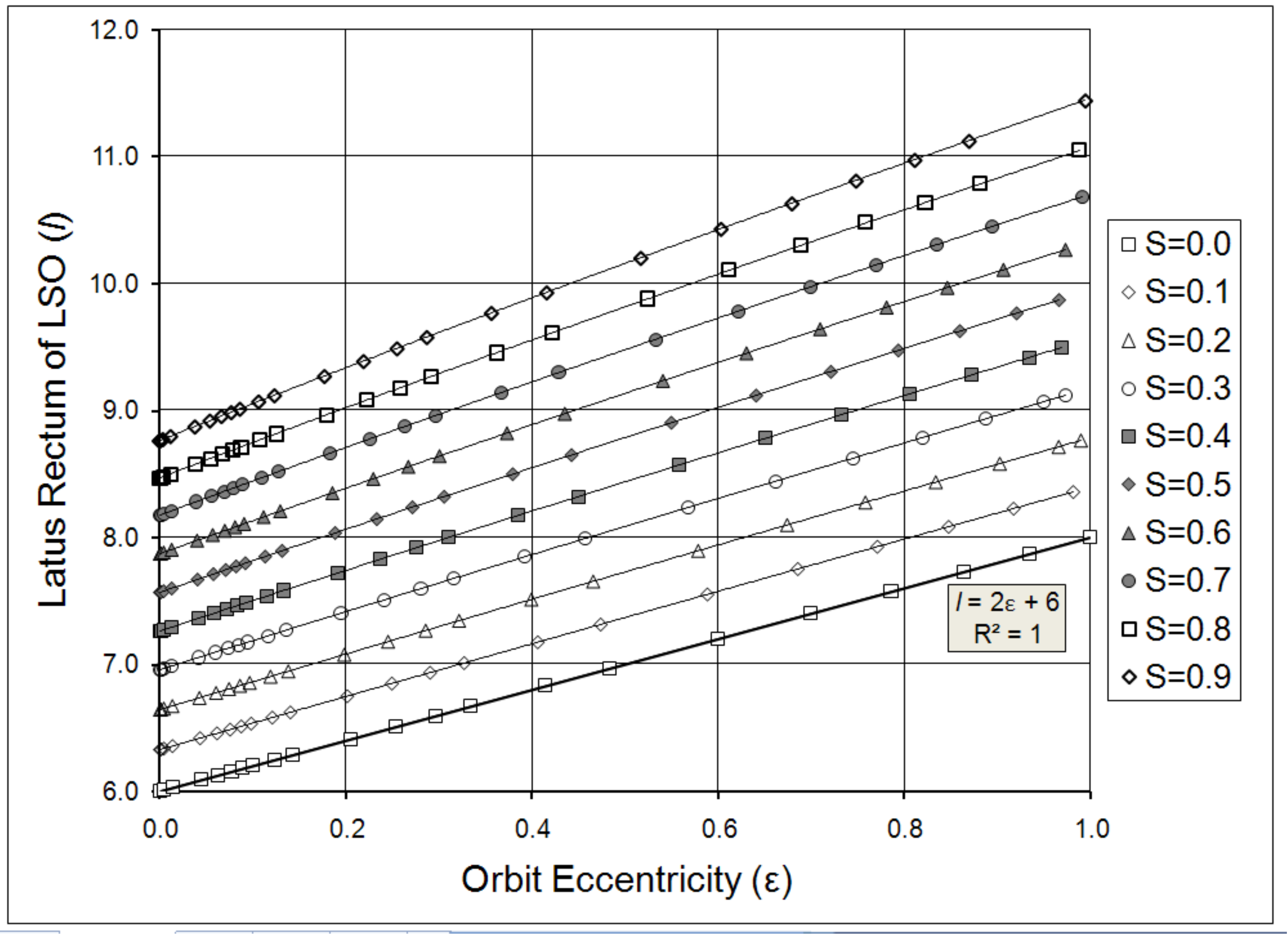}\caption{\label{fig:LSO-Retrograde}The LSO latus rectum, ${\tilde{l}}$, calculated
for KBH systems in which the test-particle is in a retrograde orbit.}

\end{sidewaysfigure}

The LSO formulae we seek will be used in future work to calculate
the test-particle orbital frequency, ${W}$, in terms of the eccentricity
of the orbit, ${\varepsilon}$, and KBH spin, ${\tilde{S}}$. One
such relationship is already known for the SBH, i.e.

\begin{eqnarray}
{\tilde{l}} & = & 6+2{\varepsilon}\label{eq:latus_rectum_SBH}\end{eqnarray}
\cite{PhysRevD.50.3816,Barack:2004uq}; but we require additional
formulae for KBH systems of various values of spin, and for the prograde
and retrograde orbits. To this end, the algorithm outlined in Section
\ref{sub:Algorithm} was used to calculate a sequence of latus rectum
values, ${\tilde{l}}$, for LSOs of differing eccentricity, ${\varepsilon}$,
in spherical coordinates. These results are plotted in Figures \ref{fig:Prograde-LSO}
and \ref{fig:LSO-Retrograde}, for prograde and retrograde orbits
respectively, and each set was fit to a sixth order polynomial equation
of the form, ${\tilde{l}}={c}_{k}\,{\varepsilon}^{k}$, where ${c}_{k}$
corresponds to the coefficients to be calculated (see Tables \ref{tab:Results-of-linear-ret}
and \ref{tab:Results-of-linear-pro}). The result for the SBH system
is shown in each of the two figures where the least squares fit yielded
a linear result, ${\tilde{l}}=6.00+2.00\,{\varepsilon}$, which is
consistent with equation (\ref{eq:latus_rectum_SBH}). Such agreement
is noteworthy because the least squares fit, based upon results previously
known through the analytical and numerical analysis described in Section
\ref{sub:Algorithm}, corroborate the LSO relationship for the SBH.

The linear approximations obtained for the KBH systems were used to
calculate the LSO radii which are essential for determining the point
at which an inspiraling CO will plunge. Although the data point pairs,
$\left(\varepsilon,\tilde{l}\right)$, derived for a particular spin
became slightly nonlinear with increasing spin, the square of the
correlation coefficient equals ${1}$.

\begin{sidewaystable}
\caption{\label{tab:Results-of-linear-ret}Results of polynomial fit applied
to the retrograde numerical data plotted in Figure \ref{fig:LSO-Retrograde}.}

\begin{tabular}{cccccccc}
\toprule 
BH Spin ${\tilde{S}}$ & \multicolumn{2}{c}{Linear Formula for ${\tilde{l}}$} & \multicolumn{5}{c}{Higher Order Elements}\tabularnewline
\midrule
 & ${\varepsilon}^{0}$ & ${\varepsilon}^{1}$ & ${\varepsilon}^{2}$ & ${\varepsilon}^{3}$ & ${\varepsilon}^{4}$ & ${\varepsilon}^{5}$ & ${\varepsilon}^{6}$\tabularnewline
\midrule
$0.9$ (ret) & $8.763688194$ & $2.921197023$ & $0.334533101$ & $0.166029604$ & $-0.090656413$ & {\small $0.038313748$} & \tabularnewline
$0.8$ (ret) & $8.469624556$ & $2.823177796$ & $0.307860108$ & $0.153608821$ & $-0.083713859$ & $0.035301568$ & $ $\tabularnewline
$0.7$ (ret) & $8.172997404$ & $2.724303684$ & $-0.279246699$ & $0.140096895$ & $-0.076041551$ & $0.031871975$ & $ $\tabularnewline
$0.6$ (ret) & $7.873580734$ & $2.624501799$ & $-0.248523134$ & $0.125602488$ & $-0.068161649$ & $0.028629143$$ $ & $ $\tabularnewline
$0.5$ (ret) & $7.571112878$ & $2.523682803$ & $-0.215396442$ & $0.109740157$ & $-0.059474839$ & $ $$0.024985751$ & $ $\tabularnewline
$0.4$ (ret) & $7.265288032$ & $2.421744495$ & $-0.179553115$ & $0.092264028$ & $-0.049805202$ & $0.020825353$ & $ $\tabularnewline
$0.3$ (ret) & $6.955745011$ & 2.318567387 & $-0.140631846$ & $0.072989398$ & $-0.039242086$ & $0.016335337$ & $ $\tabularnewline
$0.2$ (ret) & $6.6420520040$ & $2.2140067262$ & $-0.0981479990$ & $0.0514756004$ & $-0.0274318675$ & $0.0112632939$ & $ $\tabularnewline
$0.1$ (ret) & $6.3236854510$ & $2.1078897600$ & $-0.0515364094$ & $0.0274106312$ & $-0.0145526678$ & $0.0059528458$ & $ $\tabularnewline
\emph{$0.0${*}} & $6.0$ & 2.000000399 & $-0.000004828$ & $0.000019916$ & $-0.000036936$ & $0.000031685$ & $0.001939522$\tabularnewline
\bottomrule 
\end{tabular}

\begin{raggedright}
{\footnotesize {*}Consistent with \cite{PhysRevD.50.3816}; the nonlinear
elements for ${\tilde{S}}=0.0$ were probably caused by round-off
error. }
\par\end{raggedright}
\end{sidewaystable}

\begin{sidewaystable}
\caption{\label{tab:Results-of-linear-pro}Results of polynomial fit applied
to the prograde numerical data plotted in Figure \ref{fig:Prograde-LSO}.}

\begin{tabular}{cccccccc}
\toprule 
BH Spin ${\tilde{S}}$ & \multicolumn{2}{c}{Linear Formula for ${\tilde{l}}$} & \multicolumn{5}{c}{Higher Order elements}\tabularnewline
\midrule 
 & ${\varepsilon}^{0}$ & ${\varepsilon}^{1}$ & ${\varepsilon}^{2}$ & ${\varepsilon}^{3}$ & ${\varepsilon}^{4}$ & ${\varepsilon}^{5}$ & ${\varepsilon}^{6}$\tabularnewline
\midrule
\emph{$0.0${*}} & $6.0$ & 2.000000399 & $-0.000004828$ & $0.000019916$ & $-0.000036936$ & $0.000031685$ & $0.001939522$\tabularnewline
$0.1$ & $5.6701844450$ & $1.8900661368$ & $0.0575038413$ & $-0.0317686472$ & $0.0168080125$ & $-0.0068645389$ & $ $\tabularnewline
$0.2$ & $5.3331947130$ & $1.7777446758$ & $0.1223137900$ & $-0.0688936900$ & $0.0356024620$ & $-0.0138446402$ & $ $\tabularnewline
$0.3$ & $4.9876472960$ & $1.6625673677$ & $0.1965594291$ & $-0.1141250883$ & $0.0587798567$ & $-0.0225955995$ & $0.0044631526$\tabularnewline
$0.4$ & $4.6316401970$ & $1.5439026552$ & $0.2832200268$ & $-0.1707514543$ & $0.0874921671$ & $-0.0327847030$ & $ $\tabularnewline
$0.5$ & $4.2624301070$$ $ & $1.4208293794$ & $0.3872264604$ & $-0.2455511898$ & $0.1262355513$ & $-0.0461745401$ & $ $\tabularnewline
$0.6$ & $3.8757931590$ & $1.291939602$ & $0.516967848$ & $-0.350625042$ & $0.181780713$ & $-0.063577889$ & $ $\tabularnewline
$0.7$ & $3.4645809000$ & $1.154818226$ & $0.689581456$ & $-0.516713627$ & $0.279766279$ & $-0.096357483$ & $ $\tabularnewline
$0.8$ & $3.0147269360$ & $1.0047022061$ & $0.9464897749$ & $-0.8316991247$ & $0.5056498507$ & $-0.1870071888$ & $ $\tabularnewline
$0.9$ & $2.4892766210$ & $0.8291439890$ & $1.4345599659$ & $-1.6864540901$ & $1.3351280439$ & $-0.6279979225$ & $0.1324275499$\tabularnewline
\bottomrule 
\end{tabular}

\begin{raggedright}
{\footnotesize {*}Consistent with \cite{PhysRevD.50.3816}; the nonlinear
elements for ${\tilde{S}}=0.0$ were probably caused by round-off
error. }
\par\end{raggedright}
\end{sidewaystable}

\section{\label{sec:Conclusions}Conclusions}

A knowledge of the relationship between the latus rectum, ${\tilde{l}}$,
of a last stable orbit (LSO) and the Kerr black hole (KBH) spin, ${\tilde{S}}$,
where ${\tilde{S}}=\left|\mathbf{J}\right|/{M}^{2}$, is essential
for the calculation of the compact object (CO) orbit evolution in
extreme black hole systems, and thus the gravitational wave energy
emission. The Kerr metric provides the basis for the derivation of
analytical relationships between orbital angular momentum squared
(${\tilde{L}}^{2}$) and the apogee of the last stable orbit (${R_{Min}}$)
for the prograde and retrograde elliptical orbits of test-particles
about a KBH. These formulae lead directly to new and simplified representations
of ${R_{LSO}}$ with respect to KBH spin for circular orbits, which
in turn are used as a starting point in performing numerical analysis
of elliptical LSOs. By using the prograde and retrograde relationships
between the values of ${\tilde{L}}^{2}$ and ${R_{Min}}$ that we
have derived from the effective potential, an elliptical LSO can be
analysed numerically to yield values for ${R_{Min}}$. The algorithm
provides a foundation that will be generalised to include inclined
orbits for which the effective potential is more complicated than
that for the case where the orbit lies in the equatorial plane of
the KBH. Therefore finding the relationship between ${R}_{Min}$ and
orbital angular momentum becomes paramount as it allow for the methodical
treatment of orbits of successively greater eccentricity and orbital
angular momentum.

Formulae for orbital energy, ${\tilde{E}}$, and the quantity, (${\tilde{L}}-{\tilde{S}}{\tilde{E}}$),
have led us to the derivation of analytical expressions for ${\tilde{l}}$
in terms of ${\tilde{S}}$ and orbit eccentricity. The LSO values
obtained by using these formulae were in excellent agreement with
those in the literature, therefore, demonstrating their validity.
The usefulness of these analytical expressions may be found in the
advantage gained in future theoretical and numerical investigations.
These equations and the others we have derived here also demonstrate
the importance of using parameters that are normalised with respect
to the KBH mass, ${M}$.

The values of ${R}_{Min}$ and ${R_{Max}}$, in Boyer-Lindquist (BL)
coordinates, must be transformed to spherical coordinates. The ${\tilde{l}_{LSO(spherical)}}$
and LSO eccentricity (${\varepsilon}_{Spherical}$) can then be estimated
and used in the integration of the post Newtonian orbital evolution
equations. Because we can now calculate analytically the ${\varepsilon}_{Spherical}$
and ${\tilde{l}_{LSO(spherical)}}$ values for a range of KBH spins,
($0\leq{\tilde{S}}<1$; retrograde and prograde), it would facilitate
the modelling of CO orbit evolution about a massive KBH. 

The companion matrix (CM) has been shown to be of great use in finding
the roots of complicated polynomials in an analytical form. The use
of the CM in numerical work is also encouraging, especially because
one can perform various linear operations on the CM in order to transform
the final results.

Further investigation will be performed using the results of this
work as a foundation. The methodologies that underlie our numerical
algorithm will be extended to the case of inclined orbits. The radial
frequency behaviour will also be treated by performing analytical
integration of the radial path of the test-particle between the orbit
pericentre and apocentre. The post Newtonian evolution equations that
describe the inspiral of COs in extreme binary black hole systems
will then be modelled.

The authors express their gratitude to Dr. L. Rezzolla for hosting
P.G.K. at the Einstein Institute in Potsdam, Germany. P.G.K. thanks
Western Science (UWO) for their financial support. M.H.'s research
is funded through the NSERC Discovery Grant, Canada Research Chair,
Canada Foundation for Innovation, Ontario Innovation Trust, and Western's
Academic Development Fund programs. S.R.V. acknowledges research funding
by NSERC. The authors thankfully acknowledge Drs. N. Kiriushcheva
and S. V. Kuzmin for their helpful discussions of the manuscript and
I. Haque for help with the companion matrix method.

\newpage{}
\bibliographystyle{unsrt}

\section*{\newpage{}Appendix A: The Inverse Kerr Metric}

The inverse Kerr metric expressed in the Boyer-Lindquist coordinates
system. To simplify the presentation of the metric, we define the
parameter:

\begin{eqnarray}
{\Sigma}={\rho}^{2} & = & {M^{2}}{\left(R^{2}+\tilde{S}^{2}{\cos^{2}\left(\theta\right)}\right)}.\end{eqnarray}

The inverse Kerr metric is:

{\small \begin{eqnarray}
{g}^{\delta\gamma}\Biggr|_{Kerr}\label{eq:InverseKerrMetric}\\
=\left({\Sigma}\right)^{-1} & \left[{\begin{array}{cccc}
{-\frac{{{\Sigma}\left(R^{2}+\tilde{S}^{2}\right)+2\,\tilde{S}^{2}R-2\,{\cos^{2}\left(\theta\right)}\tilde{S}^{2}R}}{{\left({R^{2}-2\, R+\tilde{S}^{2}}\right)}}} & 0 & 0 & \frac{-2\,{\tilde{S}R}}{{M}{\left({R^{2}-2\, R+\tilde{S}^{2}}\right)}}\\
0 & {R^{2}-2\, R+\tilde{S}^{2}} & 0 & 0\\
0 & 0 & 1 & 0\\
\frac{-2\,{\tilde{S}R}}{{M}{\left({R^{2}-2\, R+\tilde{S}^{2}}\right)}} & 0 & 0 & {\frac{{R^{2}-2\, R+\tilde{S}^{2}{\cos^{2}\left(\theta\right)}}}{{\left({R^{2}-2\, R+\tilde{S}^{2}}\right){\sin^{2}\left(\theta\right)}}}}\end{array}}\right].\nonumber \end{eqnarray}
}{\small \par}

In this study, $\theta=\frac{\pi}{2}$, therefore, the inverse Kerr
metric simplifies to the form:

{\small \begin{eqnarray}
{g}^{\delta\gamma}{\Biggr|}_{{\normalcolor Kerr}}\label{eq:InverseKerrMetricThetaPiby2}\\
= & \left[\begin{array}{cccc}
-{\frac{{R}^{4}+{R}^{2}{\tilde{S}}^{2}+2\,{\tilde{S}}^{2}R}{\left({R}^{2}-2\, R+{\tilde{S}}^{2}\right){R}^{2}}} & 0 & 0 & -2\,{\frac{\tilde{S}}{{M}\left({R}^{2}-2\, R+{\tilde{S}}^{2}\right)R}}\\
\noalign{\medskip}0 & {\frac{{R}^{2}-2\, R+{\tilde{S}}^{2}}{{R}^{2}}} & 0 & 0\\
\noalign{\medskip}0 & 0 & \frac{1}{{M}^{2}{R}^{2}} & 0\\
\noalign{\medskip}-2\,{\frac{\tilde{S}}{{M}\left({R}^{2}-2\, R+{\tilde{S}}^{2}\right)R}} & 0 & 0 & {\frac{{R}^{2}-2\, R}{\left({R}^{2}-2\, R+{\tilde{S}}^{2}\right){M}^{2}{R}^{2}}}\end{array}\right].\nonumber \end{eqnarray}
}The determinant of the Kerr metric was calculated to be, $Det=-{\Sigma}^{2}{\sin^{2}\left(\theta\right)}.$

\newpage{}

\section*{Appendix B: Use of the Companion Matrix to Solve a Quartic Equation}

Given the task of finding the roots of a polynomial, (${p\left(R\right)=0}$),
one might proceed by regarding it to be the characteristic polynomial
of a matrix for which the eigenvalues are sought (i.e. the companion
matrix) (see chapter 7 in \cite{1996maco.book.....G}).

\begin{eqnarray}
p\left(R\right) & = & \left({R}^{4}-12\,{R}^{3}-6\,{\tilde{S}}^{2}{R}^{2}+36\,{R}^{2}-28\,{\tilde{S}}^{2}R+9\,{\tilde{S}}^{4}\right)\nonumber \\
 & = & 0.\label{eq:Characteristic-Polynomial}\end{eqnarray}
The creation of said matrix proceeds trivially to produce the companion
matrix, ${M}$ (see section 7.4.6 in \cite{1996maco.book.....G}):

\begin{eqnarray}
M & = & \left[\begin{array}{cccc}
0 & 0 & 0 & -9\,{\tilde{S}}^{4}\\
\noalign{\medskip}1 & 0 & 0 & 28\,{\tilde{S}}^{2}\\
\noalign{\medskip}0 & 1 & 0 & 6\,{\tilde{S}}^{2}-36\\
\noalign{\medskip}0 & 0 & 1 & 12\end{array}\right],\end{eqnarray}
from which one may calculate the eigenvalues. These eigenvalues represent
the solutions of equation (\ref{eq:Characteristic-Polynomial}). There
are four solutions, which are (in simplified form):\begin{eqnarray}
{R} & = & 3\pm\sqrt{{Z}}\pm\sqrt{\frac{16{\tilde{S}}^{2}}{\sqrt{{Z}}}-{Z}+3\left(3+{\tilde{S}}^{2}\right)}\end{eqnarray}
where\begin{eqnarray*}
{Z} & = & 3+{\tilde{S}}^{2}\\
 &  & +\left(3+{\tilde{S}}\right)\left(\left(1+{\tilde{S}}\right)\left(1-{\tilde{S}}\right)^{2}\right)^{\frac{1}{3}}\\
 &  & +\left(3-{\tilde{S}}\right)\left(\left(1-{\tilde{S}}\right)\left(1+{\tilde{S}}\right)^{2}\right)^{\frac{1}{3}}\end{eqnarray*}

We know by evaluating the solutions at ${\tilde{S}}=0$ (the Schwarzschild
case) which two of the four solutions ought to be retained. They are:\begin{eqnarray}
{R} & = & 3+\sqrt{{Z}}\pm\sqrt{\frac{16{\tilde{S}}^{2}}{\sqrt{{Z}}}-{Z}+3\left(3+{\tilde{S}}^{2}\right)}.\end{eqnarray}

\newpage{}

\section*{Appendix C: Use of the Companion Matrix to Find the Analytical Solution
for ${\tilde{l}}_{LSO}$ for a General Elliptical Orbit }

Treatment of the orbital energy, ${\tilde{E}}$, and the quantity,
(${X}={\tilde{L}}-{\tilde{S}}{\tilde{E}}$), leads to an analytical
expression for the latus rectum, ${\tilde{l}}$, of the last stable
orbit (LSO) of a test-particle. An analytical form of ${\tilde{E}}$
(see \cite{2002PhRvD..66d4002G}) is:\begin{eqnarray}
{\tilde{E}} & = & \sqrt{1-\left(1-{\varepsilon}^{2}\right)\left(1-{\frac{{X}^{2}\left(1-{\varepsilon}^{2}\right)}{{\tilde{l}}^{2}}}\right){\tilde{l}}^{-1}}.\end{eqnarray}
In Appendix A of \cite{2002PhRvD..66d4002G} the term ${X}^{2}$ has
been calculated to be, 

\begin{eqnarray}
{X}^{2} & = & \frac{-{n}\mp\sqrt{d}}{2{f}},\label{eq:GandKX2}\end{eqnarray}
for which the negative sign corresponds to a prograde orbit and the
positive sign corresponds to a retrograde orbit. The functions in
equation (\ref{eq:GandKX2}) are:

\begin{eqnarray}
{f} & = & {\frac{\tilde{l}\left(\tilde{l}-3-{\varepsilon}^{2}\right)^{2}-4\,{\tilde{S}}^{2}\left(1-\varepsilon\right)^{2}\left(1+\varepsilon\right)^{2}}{{\tilde{l}}^{3}}}\end{eqnarray}
and

\begin{eqnarray}
{n} & = & -2\,{\frac{\tilde{l}\left(\tilde{l}-3-{\varepsilon}^{2}\right)+{\tilde{S}}^{2}\left(\tilde{l}+1+3\,{\varepsilon}^{2}\right)}{\tilde{l}}};\end{eqnarray}
and the discriminator (${d}={n}^{2}-4{f}{c}$):

\begin{eqnarray}
{d} & = & \frac{16{\tilde{S}}^{2}}{{\tilde{l}}^{3}}\left(\tilde{l}\left(\tilde{l}-2-2\,\varepsilon\right)+{\tilde{S}}^{2}\left(1+\varepsilon\right)^{2}\right)\left(\tilde{l}\left(\tilde{l}-2+2\,\varepsilon\right)+{\tilde{S}}^{2}\left(1-\varepsilon\right)^{2}\right)\end{eqnarray}
where\begin{eqnarray*}
{c} & = & \left({\tilde{l}}-{\tilde{S}}^{2}\right)^{2}.\end{eqnarray*}

The analytical relationship between ${\tilde{l}}_{LSO}$ of the LSO
orbit and ${\tilde{S}}$ and ${\varepsilon}$ can be found by solving
either of the following equalities: \begin{eqnarray}
{\it {X_{{\it Prograde}}^{2}}}\left(1+\varepsilon\right)\left(3-\varepsilon\right) & = & {\tilde{l}}^{2}\label{eq:X2ProgradeEquation}\end{eqnarray}
and

\begin{eqnarray}
{\it {X_{Retrograde}^{2}}}_{{\it }}\left(1+\varepsilon\right)\left(3-\varepsilon\right) & = & {\tilde{l}}^{2}.\label{eq:X2RetrogradeEquation}\end{eqnarray}
By manipulating either of the equations (\ref{eq:X2ProgradeEquation})
or (\ref{eq:X2RetrogradeEquation}) and removing the square root,
one obtains the characteristic polynomial:

\begin{eqnarray}
P\left({\tilde{l}}\right) & = & {\tilde{l}}^{4}+\left(-4\,\varepsilon-12\right){\tilde{l}}^{3}\nonumber \\
 &  & +\left(-4\,{\tilde{S}}^{2}\varepsilon+2\,{\tilde{S}}^{2}{\varepsilon}^{2}-6\,{\tilde{S}}^{2}+4\,{\varepsilon}^{2}+24\,\varepsilon+36\right){\tilde{l}}^{2}\nonumber \\
 &  & -4\,{\tilde{S}}^{2}\left(1+\varepsilon\right)\left({\varepsilon}^{2}+7\right)\tilde{l}+\left(1+\varepsilon\right)^{2}\left(-3+\varepsilon\right)^{2}{\tilde{S}}^{4}\nonumber \\
 & = & 0.\label{eq:CharacteristicPolynomialElliptical}\end{eqnarray}
Converting the characteristic polynomial (equation (\ref{eq:CharacteristicPolynomialElliptical}))
into a companion matrix yields (see section 7.4.6 in \cite{1996maco.book.....G}):\begin{eqnarray}
{M} & = & \left[\begin{array}{cccc}
0 & 0 & 0 & -{\tilde{S}}^{4}\left(1+{\varepsilon}\right)^{2}\left(3-{\varepsilon}\right)^{2}\\
\noalign{\medskip}1 & 0 & 0 & 4\,{\tilde{S}}^{2}\left(1+{\varepsilon}\right)\left({\varepsilon}^{2}+7\right)\\
\noalign{\medskip}0 & 1 & 0 & -2\,{\tilde{S}}^{2}{\varepsilon}^{2}+4\,{\varepsilon}\,{\tilde{S}}^{2}+6\,{\tilde{S}}^{2}-4\,{\varepsilon}^{2}-24\,\varepsilon-36\\
\noalign{\medskip}0 & 0 & 1 & 4\,(3+\varepsilon)\end{array}\right].\label{eq:CommonMatrixElliptical}\end{eqnarray}
The eigenvalues of equation (\ref{eq:CommonMatrixElliptical}) can
be evaluated analytically and they correspond to the roots of equation
(\ref{eq:CharacteristicPolynomialElliptical}). Two of those roots
correspond to the latus rectum (${\tilde{l}}$) of each of the prograde
and retrograde test-particle orbits. One can also substitute the KBH
spin, ${\tilde{S}}$, and the eccentricity of the orbit, ${\varepsilon}$,
into the companion matrix and then calculate its eigenvalues to numerically
calculate the values of ${\tilde{l}}$. 

The analytical form of the eigenvalues is complicated; but a factorised
form is presented here to illustrate how the solutions for ${\tilde{l}}$
were identified. The four eigenvalues, ${\lambda}_{i}$ (${i}=1..4$),
are:

\begin{eqnarray}
{\lambda}_{i} & = & \left(3+\varepsilon\right)\pm\sqrt{Z_{{o}}}\label{eq:RLSO-Eigen}\\
 & \pm & \sqrt{16\,{\frac{{\tilde{S}}^{2}\left(1+\varepsilon\right)}{\sqrt{Z_{{o}}}}}-Z_{{o}}+\left(3+\varepsilon\right)^{2}+{\tilde{S}}^{2}\left(1+\varepsilon\right)\left(3-\varepsilon\right)}.\nonumber \end{eqnarray}
By following the same reasoning as in Appendix B, we know that the
solutions sought will each correspond to $6+2{\varepsilon}$ when
${\tilde{S}}=0$. In that case, ${Z}_{o}={\left(3+\varepsilon\right)^{2}}$,
therefore, equation (\ref{eq:RLSO-Eigen}) simplifies to:\begin{eqnarray}
{\lambda}_{i} & = & \left(3+\varepsilon\right)\pm\left(3+\varepsilon\right)\pm\left(0\right).\end{eqnarray}
Therefore two of the eigenvalues, where ${\lambda}_{i}=0$, are excluded.

\newpage{}

\newpage{}

\begin{sidewaysfigure}
\includegraphics[scale=0.45]{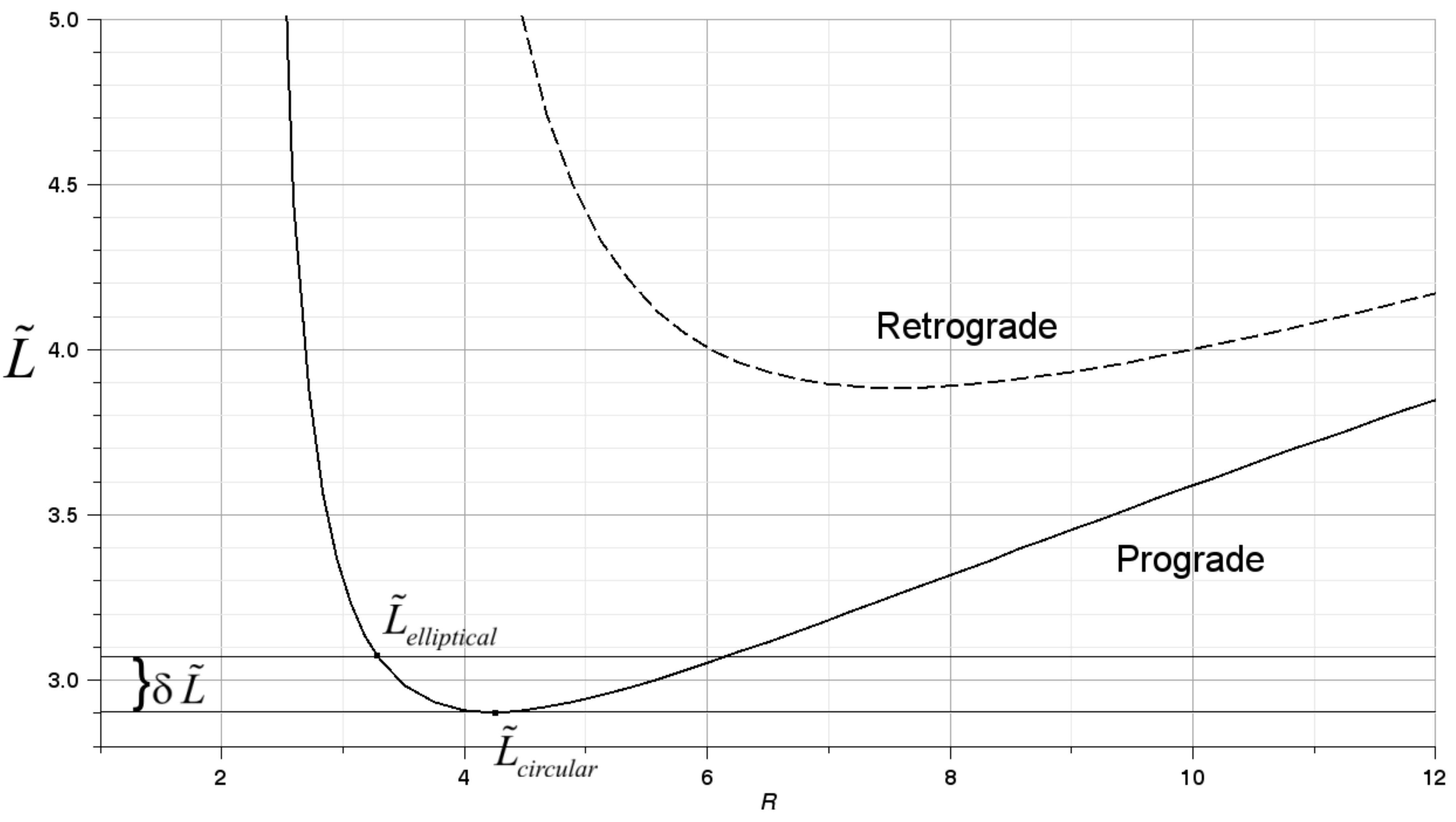}\caption{\label{fig:Min-of-L-wrt-R} The relationship between the orbital angular momentum, ${\tilde{L}}$,
and radius ${R}$ for a prograde and retrograde orbit. }

\label{fig:Min-of-L-wrt-R}
\end{sidewaysfigure}

\newpage{}

\end{document}